\documentclass[aps, pra, preprint, amsmath, amssymb, superscriptaddress]{revtex4-2}
\usepackage{graphicx}
\usepackage{dcolumn} 
\usepackage{bm}      
\usepackage{color}
\usepackage{hyperref}

\newcommand{\tr}{\mathrm{Tr\,}}
\newcommand{\dd}{\mathrm{\,d}}

\begin{document}

\title{Robust entanglement detection in arbitrary two-mode Gaussian state: a Stokes-like operator-based approach}

\author{Arijit Dutta}
\email{arijitdutta51@gmail.com}
\affiliation{Centre for Quantum Engineering, Research and Education, TCG CREST, Kolkata 700091, India}

\author{Sibasish Ghosh}
\email{sibasish@imsc.res.in}
\affiliation{Optics $\&$ Quantum Information Group, The Institute of Mathematical Sciences, HBNI, CIT Campus, Taramani, Chennai 600113, India}

\author{Jaewan Kim}
\email{jaewan@kias.re.kr}
\affiliation{Quantum Universe Center, Korea Institute for Advanced Study, Seoul 02455, Korea}

\author{Ritabrata Sengupta}
\email{rb@iiserbpr.ac.in}
\affiliation{Department of Mathematical Sciences, Indian Institute of Science Education and Research (IISER), Berhampur, Govt. ITI, Berhampur (Transit Campus), National Highway 59, Berhampur 760010, India}



\begin{abstract}
 Detection of entanglement in quantum states is one of the most important problems in quantum information processing. However, it is one of the most challenging tasks to find a {\it universal} scheme which is also desired to be optimal to detect entanglement for all states of a specific class\textendash as always preferred by experimentalists. Although, the topic is well studied at least in case of lower dimensional compound systems, e.g., two-qubit systems, but in the case of continuous variable systems, this remains as an open problem. Even in the case of two-mode Gaussian states, the problem is not fully solved. In our work, we have tried to address this issue. At first, a limited number of Hermitian operators is given to test the necessary and sufficient criterion on the covariance matrix of separable
two-mode Gaussian states. Thereafter, we present an interferometric scheme to test the same separability criterion in which the measurements are being done via Stokes-like operators. In such case, we consider only single-copy measurements on a two-mode Gaussian state at a time and the scheme amounts to the full state tomography. We further analyze the robustness of the proposed detection method against experimentally relevant imperfections and demonstrate that the separability test remains reliable under moderate levels of detection inefficiency. Although this latter approach is a linear optics based one, nevertheless it is not an economic scheme. Resource-wise a more economical scheme than
the full state tomography is obtained if we consider measurements on two copies of the state at a
time. However, optimality of the scheme is not yet known.
\end{abstract}
\maketitle
\section{Introduction}

Quantum entanglement is the fundamental aspect which
separates quantum mechanics from its classical counterpart.
Entanglement is used as resources in various 
quantum computation and information theory protocols
\cite{NC}. Hence,
detection of entanglement is important in this area.
By definition a {\em separable state} of two subsystems
$\rho_{12}$  can be written as 
\begin{equation}
\label{sepa}
\rho_{12}=\sum_i p_i\rho_1^{(i)}\otimes \rho_2^{(i)}.
\end{equation}
It  is a convex combination of product states of two
different subsystems $1$ and $2.$ The above formula
represents situation in which with probability $p_i$  one of
the system is in the state $\rho_1^{(i)}$ and  the
other is in  $\rho_2^{(j)}.$  The states
$\rho_1^{(i)},$ $\rho_2^{(i)},$ and
$\rho_{12}$ are defined on the Hilbert spaces
$\mathcal{H}_{1},$ $\mathcal{H}_{2},$ and
$\mathcal{H}_{1}\otimes \mathcal{H}_{2},$ respectively.
If a given state cannot be written in the form given by
$\eqref{sepa},$ then the state is  entangled.

\par In general, given two arbitrary Hilbert spaces $H_1$
and $H_2$, declaring whether an arbitrary quantum state
$\rho_{12}$ acting on $H_1 \otimes H_2$ is separable
or entangled is a difficult problem. In fact this problem is
  NP-hard  even for any Hilbert spaces of finite composite
dimensions greater than six \cite{Horodecki96}. However,
entanglement is used as resource in almost all quantum
information protocols. Hence, it is important to find out
methods of detection of entanglement, even for specific classes of
states only.  

\par Since the beginning of quantum information science,
continuous variable systems have also been exploited as
substitute for finite dimensional systems and possibly as
more powerful tool as well, as there is a plethora of
such research coming primarily from quantum optics, both
from theory and experiment. Most of these
continuous variable works centred around Gaussian states, as
they are easy to prepare, manipulate, and measure
\cite{RevModPhys.84.621}. Many of the  protocols in finite
dimensions have analogous protocols for continuous variables
cases as well. 
Details of applications
of continuous variables, in particular Gaussian states, can
be seen in the excellent survey article by Weedbrook et al.
\cite{RevModPhys.84.621}, and references therein.

\par In the case of continuous variables, the structure of
 states is naturally more complicated than in case of discrete systems. 
One such class of states, namely the
Gaussian states, are of particular use, as they are
relatively easy to prepare and
handle. For the Gaussian state, the central role is played by
the covariance matrix of the state. In a seminal paper~\cite{Simon00} by Simon, detection of entanglement in two-mode Gaussian state was presented where
 a continuous variable version of the Peres-Horodecki PPT criterion
 was obtained. It  had been shown that
the criterion
is necessary and sufficient to test separability of two-mode
Gaussian states.  It is worth to mention that, in Ref.~\cite{Duan00}, considering the total variance of a pair of Einstein-Podolsky-Rosen (EPR) type operators, a necessary and sufficient condition for the inseparability of any two-mode Gaussian state was proposed. Recently,  an equivalence between Simon's separability criterion and EPR-type operators based inseparability condition~\cite{Duan00} of any two-mode Gaussian state has been presented in Ref.~\cite{Marian18}.
Further generalizations of Simon's criterion~\cite{Simon00} 
are given in Refs.~\cite{Giedke03}, \cite{adesso08}, where the  criterion has
been extended for higher number of modes under certain
symmetry conditions. Tempted by the aforesaid necessary-sufficient condition of Ref.~\cite{Simon00}, one might expect that a universal entanglement witnessing scheme for (bosonic) two-mode Gaussian states should exist, like in the case of two qubits. Unfortunately, given any two-mode Gaussian state ${\rho}_{12}$, the signature of ${\rm det} ({\rho}_{12}^{T_2})$ (or, something similar to it) does not capture the necessary-sufficient condition for separability of two-mode Gaussian states, mentioned in Ref.~\cite{Simon00}. Although this necessary-sufficient condition can be cast in terms of conditions on the parameters of the covariance matrix of the two-mode Gaussian state, nevertheless, re-casting these latter conditions in terms of statistics of measurements of observables on one or more copies of the two-mode state is a non-trivial task.  

\par There are more general works in this
direction which illustrate necessary and sufficient criterion for testing separability of any Gaussian state
\cite{PhysRevLett.86.3658, krprb3}, where semi-definite
programming can be used for detection of entanglement. However, such methods can be used for computational purposes only, and cannot be used for actual experiment, where the state itself remains unknown.
An equivalent form of the separability criterion was presented in Refs.~\cite{cerf04, pirandola09}, where the separability criterion was derived in terms of the symplectic eigenvalues of the covariance matrix of a unknown two-mode Gaussian state.  An experimental friendly approach to estimate entanglement in two-mode Gaussian state was provided in Ref.~\cite{adesso04}. However,  these methods require apriori knowledge of the
covariance matrix, and as such cannot be directly used for
detection of entanglement in an unknown state. Similarly, in Ref.~\cite{cerf04}, an experimentally feasible scheme to test the separability criterion was proposed for partially known multimode Gaussian states. Precisely, in this case, the expectation values of quadrature observables are known along with the apriori knowledge whether or not the covariance matrix of the Gaussian state is symmetric. In Ref.~\cite{haruna07}, a scheme to test  a modified version of separability criterion was discussed. However, this works for two-mode Gaussian states with a special form of covariance matrix. In Ref.~\cite{Mihaescu_2020}, 
the authors have considered the question of universal detection entanglement in $k$-partite CV systems with $N$ modes (with one or more modes in possession of each part)  in which the relevant set of entanglement witness operators are being generated via semi-definite programming (SDP). These witness operators can be  realized by acting on single copy of the state at a time and using random measurements involving homodyne detections, polarizing beam spliters, and polarization rotators, a scheme which was introduced in Ref.~\cite{D'Auria05} to find out the covariance matrix of any two-mode Gaussian state (described by 14 real parameters) using five different homodyne detections ( five different field modes) in total. As this latter work actually amounts to state tomography, therefore, the scheme of Ref.~\cite{Mihaescu_2020} also amounts to state tomography, even though, one may need less (or, more) number of measurements (in comparison with the full state tomography)  for certain class of states. This is a signature of choosing the option for random measurements.

\par Given this background, in the present work, our aim is to detect entanglement
in an arbitrary two-mode Gaussian state universally with fixed set of measurement settings. For testing entanglement in two-qubit systems, there are efficient formulae. Hence,
the question is whether any such formula can also be
adopted for two-mode Gaussian states. In this paper, the question has
been answered affirmatively by using an
equivalent method of detection of entanglement in terms of
determinants of block matrices as given in Eq.~\eqref{phys31aa} (see Section~\ref{s2} for details). In this
regard, two schemes of entanglement detection have been given in which only five
particular measurements are sufficient. This is followed by implementation of the universal
scheme by performing Stokes-like measurements~\cite{PhysRevA.65.052306, PhysRevA.67.012316} on a single copy of the state at a time. Interestingly, Stokes operators are efficient tools to describe the polarization degree of freedom of states of light and phase properties of the light fields~\cite{PhysRevA.65.052306}. In addition, as the set of observables can be measured with a pair of detectors at the two outputs of an optical device, e.g., beam splitter, the scheme can be executed in a standard
experimental protocol. However, one of the drawbacks of such a scheme is that, if the photons are detected at the outputs of a measurement setup consisting of only passive linear optical devices, e.g., beam splitters and phase shifters with a fixed set of settings, then the scheme leads to the full state tomography. Thus, the scheme in Section \ref{s3} can be considered as an experimental realization of the scheme  discussed in Ref.~\cite{D'Auria05} to perform the full state tomography by homodyne detection. Interestingly, the full state tomography can be avoided by performing  a special set of measurements with SWAP operators and active nonlinear devices, e.g., optical parametric amplifiers (OPAs)~\cite{Plick-NJP-2010}  on two copies of the state at a time. 

\par The paper is organized as follows. In Section~\ref{s2}, an introductory discussion on 
Gaussian state and its entanglement detection have been given. In Section~\ref{s3}, we discuss a method to test the separability criterion $\eqref{phys31aa}$ by implementing Stokes-like measurements. However, as mentioned earlier, the scheme amounts to the full state tomography. In Section~\ref{sec4}, we present another scheme to test the separability criterion $\eqref{phys31aa},$ in which one can  skip the  full state tomography by considering measurements on two copies of the unknown Gaussian state at a time. Finally, in Section~\ref{s5}, we discuss the main results and future directions and talk about a few open problems. In Appendix~\ref{app2}, two schemes to test separability criterion \eqref{phys31aa} with only five measurements have also been described. 

 \section{Test of separability and entanglement monotone} \label{s2}
\par In our present work, we are interested in detection of
entanglement in an unknown two-mode Gaussian state. Since
the state itself is unknown, we don't have any information
about its covariance matrix, and hence one can not use Simon's
criterion~\cite{Simon00} directly. Moreover, it also requires us to perform tomography on
the state to estimate all necessary parameters. 
It is well known
fact that a Gaussian state $\rho_{12}$ can be
completely characterized  by the first and second
moments. Precisely, the
Wigner function of any two-mode Gaussian state
$\rho_{12}$ reads as
 \begin{eqnarray}
 \label{Gauss}
 W_{\rho_{12}}(\vec{\xi})=\frac{1}{4\pi^2\sqrt{\det\Gamma_{\rho_{12}}}}\exp\left\{-\frac{1}{2}\left(\vec{\xi}-\langle
	 \vec{R}\rangle\right)^T
	 \Gamma_{\rho_{12}}^{-1}\left(\vec{\xi}-\langle
	 \vec{R}\rangle\right)\right\}
 \end{eqnarray}
where $\vec{\xi}=(q_1, p_1, q_2, p_2)^T$  is a vector of
phase-space point in $\mathcal{R}^4$ and $\hat{R}=(\hat{q}_1, \hat{p}_1,
\hat{q}_2, \hat{p}_2)^T$ is a vector of phase-space observables $\hat{q}_1,
\hat{p}_1, \hat{q}_2, \hat{p}_2$ satisfying commutation relation $[\hat{R}_k,
\hat{R}_l]=iJ_{kl},$ where $J$ is a matrix defined as $J=\bigoplus^2_{i=1}
\omega$ for $\omega= \begin{pmatrix}
 0 & 1  \\ -1  &0\end{pmatrix}.$ The first moments and
 covariance matrix of the state $\rho_{12}$ are
 defined as $d_j=\langle \hat{R}_j\rangle,$ and
 \[\Gamma_{\rho_{12}}=((\Gamma_{\rho_{12}})_{k,
 l})_{k,
 l=1}^4=\left(\frac{1}{2}\langle\hat{R}_k\hat{R}_l+\hat{R}_l\hat{R}_k\rangle-\langle\hat{R}_k\rangle\langle\hat{R}_l\rangle\right)_{k,
 l=1}^4=\left(\frac{1}{2}\langle\{\hat{R}_k,
 \hat{R}_l\}\rangle-\langle\hat{R}_k\rangle\langle\hat{R}_l\rangle\right)_{k,
 l=1}^4,\]
 receptively, where $\langle \hat{O}\rangle=\tr(\rho_{12}\hat{O}).$ 

For a two-mode Gaussian state $\rho_{12},$ the following condition must hold (because of positive semi-definiteness of $\rho_{12}$)
\begin{eqnarray}
 \label{phys1}
 \Gamma_{\rho_{12}}+\frac{ i}{2}J\geq 0.
 \end{eqnarray}
 Here $\Gamma_{\rho_{12}}$ can be taken  as 
 $\begin{pmatrix}
 A & C  \\ C^T  &B
 \end{pmatrix},$ where $A, B, C$ are matrices of dimension $2\times2.$ 
It is worth to mention that, by local symplectic operations on each  single modes, one can set the first moments of the state  to zero, i.e., $\langle \hat{q}_1\rangle=\langle \hat{q}_2\rangle=\langle \hat{p}_1\rangle=\langle \hat{p}_2\rangle=0,$ which, of course requires apriori information regarding the values of these first moments of the state.
 
\par To detect entanglement in case of  a two-mode Gaussian state, we employ an equivalent version \cite{cerf04, pirandola09} of 
	Simon's  separability criterion~\cite{Simon00}, which is direct generalization of
	the Peres-Horodecki separability criterion \cite{Peres96, Horodecki96} of partial transpose to continuous variable systems. Under  partial
	transpose, Wigner distribution of the two-mode Gaussian state undergoes
	mirror reflection, which leads to the transformation in
	$\Gamma_{\rho_{12}}:$ $\Gamma_{\rho_{12}}\to
	\tilde{\Gamma}_{\rho_{12}}=\Lambda \Gamma_{\rho_{12}}
	\Lambda,$ where $\Lambda$ is phase space mirror reflection. Using
	$\eqref{phys1},$ Simon's  criterion here reads 
\begin{eqnarray}
 \label{phys2}
\tilde{\Gamma}_{\rho_{12}}+\frac{ i}{2}J\geq 0.
 \end{eqnarray}

	One can obtain an analytical formula 
	for $\xi_{\text{min}}$ by solving a bi-quadratic equation~\cite{vidal02, cerf04}:
\begin{eqnarray}
 \label{phys30}
 &\xi^4-(\det A+ \det B-2\det C)\xi^2+\det\Gamma_{\rho_{12}}=0.&
 \end{eqnarray} 
 Symplectic eigenvalues of  $\tilde{\Gamma}_{\rho_{12}}$ are two positive
 roots of $\eqref{phys30},$ and  the minimum symplectic eigenvalue 
 $\xi_{\text{min}}$ \cite{cerf04, pirandola09} reads:
 \begin{eqnarray}
 \label{phys31}
 &\xi_{\text{min}}=\sqrt{\frac{D-\sqrt{D^2-4\text{det} \Gamma_{\rho_{12}}}}{2}},&
 \end{eqnarray} 
 where $D=\det A+ \det B-2\det C.$ Inserting $\xi_{\text{min}}$ in separability criterion $(\xi_{\text{min}}\geq 1/2)$, we obtain the following necessary-sufficient criterion for separability of a two-mode Gaussian state ${\rho}_{12}$:
 \begin{eqnarray}
 \label{phys31aa}
&D-4\det \Gamma_{\rho_{12}}\leq \frac{1}{4}&
\end{eqnarray}
 In addition, there exist entanglement  monotones
	\cite{vidal02, Giedke03, Wolf04}, which measure
	entanglement in a two-mode Gaussian state
	$\rho_{12}.$ For example, one can calculate
	the logarithmic negativity $\mathbb{E}_2$ to
	quantify  entanglement in $\rho_{12}.$ The
	logarithmic negativity $\mathbb{E}_2$ is  defined as
	$\mathbb{E}_2=-\log \xi_{\text{min}}.$
	In this context, it is worth to mention that in Ref.~\cite{cerf04} a scheme has been provided to measure squeezing and detect entanglement in
	multimode Gaussian states with phase-insensitive devices without
	homodyning. 
Interestingly, the scheme doesn't provide full information of the
two-mode Gaussian state. Precisely, the estimation of the determinants of $\det
A, \,\det B,\, \det C,$ and $\det \Gamma_{\rho_{12}}$ do not lead to
the full state tomography.
However, in Ref.~\cite{cerf04}, the information about the expectation values of
quadratures is necessary to choose the measurement setups to estimate the
determinants of the block matrices. Precisely, in case of no apriori knowledge about the
expectation values of quadratures one does not seem to have the potentiality to decide if the measurements for entanglement detection should be
performed on a single copy or two copies of the two-mode Gaussian state. In
addition, the method to determine $\det C,$ requires a symmetric form of the
block matrix $C.$ If the matrix is not in a symmetric form, then a proper unitary transformation is  needed to symmetrize the matrix $C.$ Hence, apriori knowledge
on the symmetry of matrix $C$ is also necessary. Therefore, the
method described in Ref.~\cite{cerf04} does not address the issue of universal
entanglement detection.
Similarly, in Ref.~\cite{haruna07}, the authors proposed a scheme to test and quantify entanglement in an arbitrary two-mode Gaussian state with minimal requirements via local measurements and a classical communication channel. To this end, the authors developed a method to test a variant  of separability criterion $\eqref{phys31aa}$ and estimate entanglement of formation to quantify entanglement of the Gaussian state. However, the method was discussed for a specific form of the covariance matrix associated to a two-mode Gaussian state. Note that, a covariance matrix of an arbitrary two-mode Gaussian state can be transformed to this specific form by local symplectic transformations, but, to get such form one must have knowledge on the elements of the covariance matrix before the transformation is being applied upon. Apart from that, the number of measurements to execute such a scheme is more than the number of measurements required to perform the  full state tomography. 
In the present work, our main  aim is to provide schemes to test the separability criterion~\eqref{phys31aa} without the requirement of  any apriori knowledge of the unknown two-mode Gaussian state. For measurements on a single copy of the state at a time, the scheme leads to the full state tomography, and the number of measurements is  same as that required for the  full state tomography  using projective measurement. However, for a special set of measurements on two copies of the state at a time, one doesn't need to find each elements of the covariance matrix of an arbitrary two-mode Gaussian state. In other words, this latter scheme is not identical to the full state tomography. 
\par 


  \begin{figure}[ht!]
  \includegraphics[trim=250 275 250 300 width=1.75\linewidth]{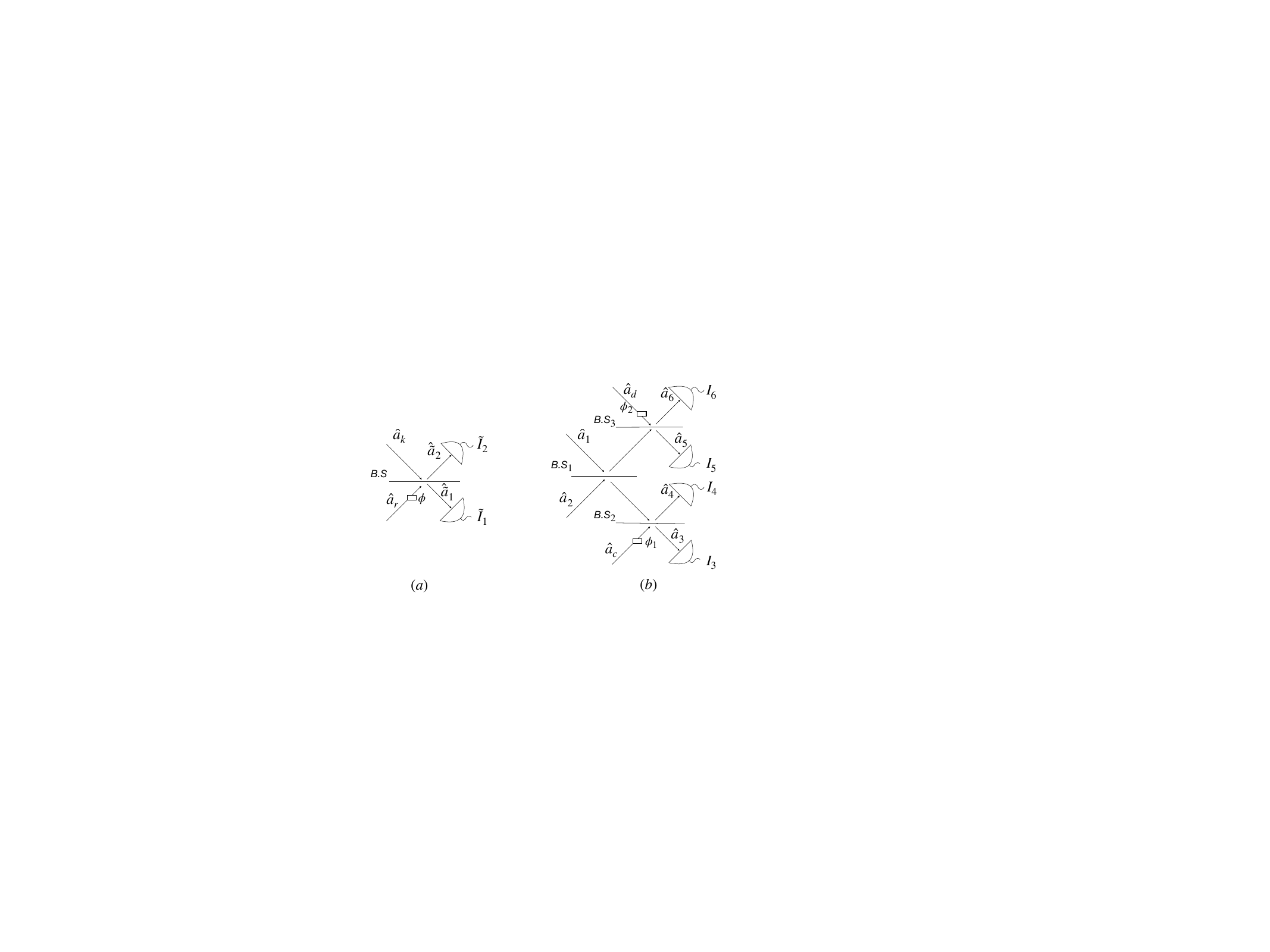}
\caption{a) A schematic diagram of the measurement setup to estimate all elements of a covariance matrix associated to a single mode of a Gaussian state. k-th mode of an unknown Gaussian state and a reference state which are represented by annihilation operators $\hat{a}_k,$ and $\hat{a}_r,$ respectively, interfere at a 50-50 beam-splitter ($\text{B.S}).$ A phase shifter is introduced at the reference mode. The outputs of the beam-splitter are represented by annihilation operators $\hat{\tilde{a}}_1,$ and $\hat{\tilde{a}}_2.$ The detectors generate photocurrents $\tilde{I}_1$ and $\tilde{I}_2$ which are proportional to the intensities at the output modes of \text{B.S}.
b) A schematic diagram of the measurement setup to obtain all elements of a covariance matrix corresponding to a two-mode Gaussian state. Two modes of $\rho_{12}$
 represented by annihilation operators $\hat{a}_1,$ and $\hat{a}_2,$ interfere at
	  a 50-50 beam-splitter 1 ($\text{B.S}_1).$ The outputs of $\text{B.S}_1$
	  are made to interfere at two separate 50-50 beam-slitters, e.g.,
	  beam-splitter 2 ($\text{B.S}_2$) and  beam-splitter 3 ($\text{B.S}_3$)
	  along with the two reference states represented by annihilation
	  operators $\hat{a}_c,$ and $\hat{a}_d.$ Phase shifters are introduced
	  at the reference modes. The outputs of the measurement setup are
	  represented by annihilation operators $\hat{a}_3,$ $\hat{a}_4,$
	  $\hat{a}_5,$ and $\hat{a}_6.$ Two pairs of photocurrents $(I_3, I_4),$
	  and $(I_5, I_6)$ are proportional to the intensities of the output
	  modes of $\text{B.S}_2$ and $\text{B.S}_3,$ respectively.
	}
\label{figg1}
\end{figure}

 \section{Measurements (Experimentally feasible)}\label{s3}
As discussed before, to test the  separability criterion
$\eqref{phys31aa},$ one needs to compute  determinants of the block matrices $A,
B,$ and $C$ of the covariance matrix of a two-mode Gaussian state
$\rho_{12}$ together with $\det {\Gamma_{\rho_{12}}}.$ One of the methods is to estimate all the elements of the
$\Gamma_{\rho_{12}}$ and calculate the determinants. Here, we present such
a scheme which can be realized in linear optical setups. Note that, as the
two-mode Gaussian state is unknown, we don't assume any properties of the
Gaussian state to perform the measurements. Hence, a general method is adopted
such that it works for an arbitrary two-mode Gaussian state.

Interestingly, in Ref.~\cite{Ruppert16}, a similar approach was implemented for a single-mode Gaussian state. 
To this end, the authors proposed a scheme in which a reference Gaussian state is made to interfere with the unknown single-mode Gaussian state on a beam-splitter and Stokes-like measurements are performed by measuring intensity difference  at the two output modes of the beam-splitter. The experimental data along with the known values of means, variances and expectation value of a symmetric function of a pair of quadrature observables related to the reference mode are sufficient to estimate the means, variances and expectation value of the symmetric function of the pair of quadrature observables associated to the unknown single-mode Gaussian state. 
We consider the same method to estimate all the elements of matrices $A$ and $B.$ 
 To obtain the elements of matrix $C$, we extend the scheme by modifying the experimental setup with additional beam-splitters and phase-shifters. The details of the such scheme with Stokes-like measurements are given below.

Let's consider $\hat{a}_k$ be the annihilation operator (see Figure~\ref{figg1}(a))
associated to k-th mode of a single-mode Gaussian state and an annihilation
operator $\hat{a}_r$ represents a reference mode $r$. A phase-shifter is placed
at the input mode of the 50-50 beam-splitter corresponding to the reference state
to introduce  phase shifts between two interfering modes. After relevant unitary
transformations by a combination of a phase shifter and a beam splitter,
annihilation operators $\hat{\tilde{a}}_1$ and $\hat{\tilde{a}}_2$ of the output
modes of the beam-splitter are given as follows:
\begin{eqnarray}
\label{single}
\hat{\tilde{a}}_1=(\hat{a}_k-\hat{a}_re^{i\phi})/\sqrt{2},~~\text{and}~~
\hat{\tilde{a}}_2=(\hat{a}_k+\hat{a}_re^{i\phi})/\sqrt{2}.
\end{eqnarray}
Also, quadrature observables  $\hat{q}_k,$ and $\hat{p}_k$ associated to k-th mode are defined as $(\hat{a}_k+\hat{a}_k^{\dagger})/\sqrt{2},$ and  $i(\hat{a}^{\dagger}_k-\hat{a}_k)/\sqrt{2},$ respectively. Throughout the
manuscript we follow the similar convention. Note that, usually Stokes operators $\hat{S}_0, \hat{S}_1,\hat{S}_2,$ and $\hat{S}_3$ are associated to two orthogonal polarization modes of a single beam~\cite{PhysRevA.65.052306}. However, Stokes-like measurements  can be realized on two output modes of a beam-splitter where the interfering input beams are generated from two different sources. To estimate the elements of matrices $A,$ and $B,$ which are associated to single modes,  $\hat{S}_1(\phi),$ $\hat{S}^2_1(\phi)$  are measured for different  phase shifts $\phi$.

Considering photon number difference at the two outputs, we obtain 
\begin{eqnarray}
\label{stokes00}
\hat{S}_1(\phi)=\tilde{I}_{2}-\tilde{I}_{1}=\hat{\tilde{a}}_{2}^{\dagger}\hat{\tilde{a}}_{2}-\hat{\tilde{a}}_{1}^{\dagger}\hat{\tilde{a}}_{1}=\hat{q}_k\hat{q}_r^{\phi}+\hat{p}_k\hat{p}_r^{\phi},
 \end{eqnarray}
where $\hat{q}_r^{\phi}=\hat{q}_r\cos\phi-\hat{p}_r\sin\phi,$ and $\hat{p}_r^{\phi}=\hat{q}_r\sin\phi+\hat{p}_r\cos\phi.$ Averaging over many copies of the k-th mode of the Gaussian state, one can estimate the expectation value of $\hat{S}_1(\phi),$ i.e.,
\begin{eqnarray}
\label{stokes01}
\langle\hat{S}_1(\phi)\rangle=\langle\hat{q}_k\rangle\langle\hat{q}_r^{\phi}\rangle+\langle\hat{p}_k\rangle\langle\hat{p}_r^{\phi}\rangle.
 \end{eqnarray}
Note that, the expectation values of quadrature observables associated to reference state are given (Here we consider a displaced squeezed thermal state as a reference state. Please see Appendix~\ref{reference} for the details). Thus, one can estimate $\langle\hat{q}_k\rangle$ and $\langle\hat{p}_k\rangle$ for experimentally obtained expectation values of $\hat{S}_1(\phi),$ and  given values of $(\langle\hat{q}_r^{\phi}\rangle$, $\langle\hat{p}_r^{\phi}\rangle)$ by solving two linear equations obtained from $\eqref{stokes01}$ for two different phase shifts, e.g., $\phi=0, \frac{\pi}{2}.$

Thereafter, estimating $\langle\hat{q}^2_k\rangle$ and $\langle\hat{p}^2_k\rangle$  we can compute variances of quadrature observables. To this end, we calculate expectation values of the  square of the difference of photon counts at the two output modes of the beam-splitter for two different values of $\phi.$ Precisely, the expectation values read
\begin{equation}
\label{stokes02}
\langle\hat{S}_1^2(\phi=0)\rangle=\langle(\tilde{I}_{2}-\tilde{I}_{1})^2\rangle_{\phi=0}=\langle\hat{q}^2_k\rangle\langle\hat{q}^2_r\rangle+\langle\hat{p}^2_k\rangle\langle\hat{p}^2_r\rangle
+\langle\hat{q}_k\hat{p}_k\rangle\langle\hat{q}_r\hat{p}_r\rangle+\langle\hat{p}_k\hat{q}_k\rangle\langle\hat{p}_r\hat{q}_r\rangle,
 \end{equation}
 and
\begin{equation}
\label{stokes03}
\langle\hat{S}^2_1(\phi=\frac{\pi}{2})\rangle=\langle(\tilde{I}_{2}-\tilde{I}_{1})^2\rangle_{\phi=\frac{\pi}{2}}=\langle\hat{q}^2_k\rangle\langle\hat{p}^2_r\rangle+\langle\hat{p}^2_k\rangle\langle\hat{q}^2_r\rangle
-\langle\hat{q}_k\hat{p}_k\rangle\langle\hat{p}_r\hat{q}_r\rangle-\langle\hat{p}_k\hat{q}_k\rangle\langle\hat{q}_r\hat{p}_r\rangle.
 \end{equation}
For proper choice of the reference state one can set $\langle\hat{q}_r\hat{p}_r\rangle=-\langle\hat{p}_r\hat{q}_r\rangle=i/2$ (please see Eq.~$\eqref{append6}$). Thereafter, solving $\eqref{stokes02}$, and $\eqref{stokes03},$ we obtain $\langle\hat{q}^2_k\rangle$ and $\langle\hat{p}^2_k\rangle.$ In Appendix~\ref{reference}, we compute $\langle\hat{q}^2_r\rangle$ and $\langle\hat{p}^2_r\rangle$  for a known displaced squeezed thermal state. 
To estimate expectation value of the symmetric function of $\hat{q}_k$ and $\hat{p}_k,$ one needs to experimentally obtain the expectation value of $\hat{S}_1^2(\phi=\frac{\pi}{4}).$ Note that, using $\eqref{stokes00},$ $\langle\hat{S}^2_1(\phi=\frac{\pi}{4})\rangle$ can be expressed as follows
  \begin{eqnarray}
\label{stokes033}
&\left\langle\hat{S}^2_1\left(\phi=\frac{\pi}{4}\right)\right\rangle = 
\left\langle\left(\tilde{I}_{2}-\tilde{I}_{1}\right)^2\right\rangle_{\phi=\frac{\pi}{4}} &\nonumber\\&
= \frac{1}{ 2}\big( \left\langle\hat{q}^2_k+\hat{p}^2_k \right\rangle \left\langle\hat{q}^2_r+\hat{p}^2_r\right\rangle - \left\langle\hat{q}^2_k-\hat{p}^2_k \right\rangle \left\langle\hat{q}_r\hat{p}_r+\hat{p}_r\hat{q}_r \right\rangle  +\left\langle\hat{q}_k\hat{p}_k+\hat{p}_k\hat{q}_k\right\rangle \left\langle\hat{q}^2_r-\hat{p}^2_r \right\rangle-1\big),
 \end{eqnarray}
 
 where we use $\left[\hat{q}_k, \hat{p}_k\right]=\left[\hat{q}_r,
 \hat{p}_r\right]=i\openone.$ Thereafter, inserting  the values of
 $\langle\hat{q}^2_k\rangle$, $\langle\hat{p}^2_k\rangle,$
 $\langle\hat{q}^2_r\rangle$, $\langle\hat{p}^2_r\rangle,$ and
 $\langle\hat{q}_r\hat{p}_r+\hat{p}_r\hat{q}_r\rangle$ in $\eqref{stokes03},$ we
 can find out $\langle\hat{q}_k\hat{p}_k+\hat{p}_k\hat{q}_k\rangle.$

In Figure~\ref{figg1}(b), we present a schematic diagram of Stokes-like
measurements. It is shown below that three $50$-$50$ beam-splitters and two 
phase shifters $(\phi_1, \phi_2)$ are sufficient to estimate all the elements of $C.$ Here $\hat{a}_1$
and $\hat{a}_2$ are annihilation operators  associated to the two modes of the
Gaussian state $\rho_{12}.$ Similarly, $\hat{a}_c$ and $\hat{a}_d$ are
annihilation operators of the two modes of a pair of single-mode reference states. Here,
$\hat{a}_1,$ $\hat{a}_2,$ $\hat{a}_c,$ and $\hat{a}_d$ are the annihilation
operators of input modes of the measurement setup, whereas  $\hat{a}_3,$
$\hat{a}_4,$ $\hat{a}_5,$ and $\hat{a}_6$ are the annihilation operators
associated to output modes. Phase shifters at the input modes of the reference
states can be changed by an observer. Here, the elements of the matrix $C$ are computed by measuring  $\hat{S}_1({\phi}_1)$, $\hat{S}_1({\phi}_2),$ $\hat{S}_1^2({\phi}_1)$, $\hat{S}_1^2({\phi}_2),$ and $\hat{S}_3(\phi_1, \phi_2).$
The intensity difference at the two output
modes of beam-splitter 2 is computed as follows
\begin{eqnarray}
\label{stokes1}
&\hat{S}_1(\phi_1)=I_{4}-I_{3}=\hat{a}_4^{\dagger}\hat{a}_4-\hat{a}_3^{\dagger}\hat{a}_3&\nonumber\\&=\frac{1}{\sqrt{2}}((\hat{q}_1-\hat{q}_2)\hat{q}_c^{\phi_1}+(\hat{p}_1-\hat{p}_2)\hat{p}_c^{\phi_1}).&
 \end{eqnarray}
 Similarly,  the intensity difference at the two output modes of beam-splitter 3 reads
\begin{eqnarray}
 \label{stokes2}
&\hat{S}_1(\phi_2)=I_{6}-I_{5}=\hat{a}_6^{\dagger}\hat{a}_6-\hat{a}_5^{\dagger}\hat{a}_5&\nonumber\\&=\frac{1}{\sqrt{2}}((\hat{q}_1+\hat{q}_2)\hat{q}_d^{\phi_2}+(\hat{p}_1+\hat{p}_2)\hat{p}_d^{\phi_2}).&
 \end{eqnarray}
 
Another type of Stokes-like measurement associated to joint measurement on the two output modes of beam-splitters 2 and 3 is given as
\begin{eqnarray}
\label{stokes5}
&\hat{S}_3(\phi_1, \phi_2)=\imath(\hat{a}_6^{\dagger}\hat{a}_3-\hat{a}_3^{\dagger}\hat{a}_6)=\frac{1}{2\sqrt{2}}((\hat{q}_1-\hat{q}_2)\hat{p}_d^{\phi_2}+(\hat{q}_1+\hat{q}_2)\hat{p}_c^{\phi_1}-(\hat{p}_1-\hat{p}_2)\hat{q}_d^{\phi_2}-(\hat{p}_1+\hat{p}_2)\hat{q}_c^{\phi_1})&\nonumber\\&
+\frac{1}{2}(\hat{q}_1\otimes\hat{p}_2-\hat{p}_1\otimes\hat{q}_2)+\frac{1}{2}(\hat{q}_c\otimes\hat{q}_d \sin(\phi_1-\phi_2)&\nonumber\\&+\hat{p}_c\otimes\hat{p}_d \sin(\phi_1-\phi_2)+\hat{q}_d\otimes\hat{p}_c \cos(\phi_1-\phi_2)-\hat{q}_c\otimes\hat{p}_d \cos(\phi_1-\phi_2)).&
\end{eqnarray}
Here, anti-coincidence of the detected photons is to be detected by using photon
number resolving detectors at the two output modes.

After performing
measurements on many copies of the state
$\rho_{12}\otimes\rho_{c}\otimes\rho_{d}$ one can estimate the
expectation values of $\hat{S}_1^2(\phi_1=0), \hat{S}_1^{2}(\phi_2=0),
\hat{S}_1^{2}(\phi_2=\frac{\pi}{4}), \hat{S}_1(\phi_1=0)\otimes\hat{S}_1(\phi_2=0),$ and  $\hat{S}_3(\phi_1=0, \phi_2=0).$ 
After, simplification we can write
\begin{eqnarray}
\label{stokes6}
&\langle\hat{S}_1^{2}(\phi_1=0)\rangle=\langle(I_{4}-I_{3})^2\rangle=\langle(\hat{a}_4^{\dagger}\hat{a}_4-\hat{a}_3^{\dagger}\hat{a}_3)^2\rangle&\nonumber\\&
=\frac{1}{2}((\langle\hat{q}_1^2\rangle-2\langle\hat{q}_1\otimes\hat{q}_2\rangle+\langle\hat{q}_2^2\rangle)\langle\hat{q}_c^2\rangle+(\langle\hat{q}_1\hat{p}_1\rangle-\langle\hat{q}_1\otimes\hat{p}_2\rangle-\langle\hat{p}_1\otimes\hat{q}_2\rangle+\langle\hat{q}_2\hat{p}_2\rangle)\langle\hat{q}_c\hat{p}_c\rangle&\nonumber\\&+(\langle\hat{p}_1\hat{q}_1\rangle-\langle\hat{q}_1\otimes\hat{p}_2\rangle-\langle\hat{p}_1\otimes\hat{q}_2\rangle+\langle\hat{p}_2\hat{q}_2\rangle)\langle\hat{p}_c\hat{q}_c\rangle+(\langle\hat{p}_1^2\rangle-2\langle\hat{p}_1\otimes\hat{p}_2\rangle+\langle\hat{p}_2^2\rangle)\langle\hat{p}_c^2\rangle)),
 \end{eqnarray}

 \begin{eqnarray}
\label{stokes7}
&\langle\hat{S}_1^{2}(\phi_2=0)\rangle=\langle(I_{6}-I_{5})^2\rangle=\langle(\hat{a}_6^{\dagger}\hat{a}_6-\hat{a}_5^{\dagger}\hat{a}_5)^2\rangle
&\nonumber\\&
=\frac{1}{2}((\langle\hat{q}_1^2\rangle+2\langle\hat{q}_1\otimes\hat{q}_2\rangle +\langle\hat{q}_2^2\rangle)\langle\hat{q}_d^2\rangle+(\langle\hat{q}_1\hat{p}_1\rangle+\langle\hat{q}_1\otimes\hat{p}_2\rangle+\langle\hat{p}_1\otimes\hat{q}_2\rangle+\langle\hat{q}_2\hat{p}_2\rangle)\langle\hat{q}_d\hat{p}_d\rangle&\nonumber\\&+(\langle\hat{p}_1\hat{q}_1\rangle+\langle\hat{q}_1\otimes\hat{p}_2\rangle+\langle\hat{p}_1\otimes\hat{q}_2\rangle+\langle\hat{p}_2\hat{q}_2\rangle)\langle\hat{p}_d\hat{q}_d\rangle+(\langle\hat{p}_1^2\rangle+2\langle\hat{p}_1\otimes\hat{p}_2\rangle+\langle\hat{p}_2^2\rangle)\langle\hat{p}_d^2\rangle)),
\end{eqnarray}
\begin{eqnarray}
\label{stokes8}
&\langle\hat{S}_1^{2}(\phi_2=\frac{\pi}{4})\rangle=\frac{1}{4}\langle((\hat{q}_1
+\hat{q}_2)(\hat{q}_d-\hat{p}_d)+(\hat{p}_1
+\hat{p}_2)(\hat{q}_d+\hat{p}_d))^2\rangle&\nonumber\\&=\frac{1}{4}((\langle \hat{q}_1^2\rangle+2\langle\hat{q}_1\otimes\hat{q}_2\rangle+\langle\hat{q}_2^2\rangle)(\langle \hat{q}_d^2\rangle-\langle\hat{q}_d\hat{p}_d\rangle-\langle\hat{p}_d\hat{q}_d\rangle+\langle\hat{p}_d^2\rangle)&\nonumber\\&+(\langle \hat{p}_1^2\rangle+2\langle\hat{p}_1\otimes\hat{p}_2\rangle+\langle\hat{p}_2^2\rangle)(\langle \hat{q}_d^2\rangle+\langle\hat{q}_d\hat{p}_d\rangle+\langle\hat{p}_d\hat{q}_d\rangle+\langle\hat{p}_d^2\rangle)&\nonumber\\&+(\langle \hat{q}_1\hat{p}_1\rangle+\langle\hat{q}_1\otimes\hat{p}_2\rangle+\langle\hat{p}_1\otimes\hat{q}_2\rangle+\langle\hat{q}_2\hat{p}_2\rangle)(\langle \hat{q}_d^2\rangle+\langle\hat{q}_d\hat{p}_d\rangle-\langle\hat{p}_d\hat{q}_d\rangle-\langle\hat{p}_d^2\rangle)&\nonumber\\&+(\langle \hat{p}_1\hat{q}_1\rangle+\langle\hat{p}_1\otimes\hat{q}_2\rangle+\langle\hat{q}_1\otimes\hat{p}_2\rangle+\langle\hat{p}_2\hat{q}_2\rangle)(\langle \hat{q}_d^2\rangle-\langle\hat{q}_d\hat{p}_d\rangle+\langle\hat{p}_d\hat{q}_d\rangle-\langle\hat{p}_d^2\rangle)),&\end{eqnarray}

\begin{eqnarray}
\label{stokes10}
&\langle\hat{S}_1(\phi_1=0)\otimes \hat{S}_1(\phi_2=0)\rangle=\frac{1}{2}((\langle\hat{q}_1^2\rangle-\langle\hat{q}_2^2\rangle)\langle\hat{q}_c\rangle\langle\hat{q}_d\rangle+(\langle\hat{p}_1^2\rangle-\langle\hat{p}_2^2\rangle)\langle\hat{p}_c\rangle\langle\hat{p}_d\rangle&\nonumber\\&+(\langle\hat{q}_1\hat{p}_1\rangle-\langle\hat{q}_2\hat{p}_2\rangle)\langle\hat{q}_c\rangle\langle\hat{p}_d\rangle+(\langle\hat{p}_1\hat{q}_1\rangle-\langle\hat{p}_2\hat{q}_2\rangle)\langle\hat{p}_c\rangle\langle\hat{q}_d\rangle+(\langle\hat{q}_1\otimes\hat{p}_2\rangle-\langle\hat{p}_1\otimes\hat{q}_2\rangle)(\langle\hat{q}_c\rangle\langle\hat{p}_d\rangle-\langle\hat{p}_c\rangle\langle\hat{q}_d\rangle)),&
\end{eqnarray}
and
\begin{eqnarray}
\label{stokes9}
&\langle\hat{S}_3(\phi_1=0, \phi_2=0)\rangle=\frac{1}{2\sqrt{2}}((\langle\hat{q}_1\rangle-\langle\hat{q}_2\rangle)\langle\hat{p}_d\rangle+(\langle\hat{q}_1\rangle+\langle\hat{q}_2\rangle)\langle\hat{p}_c\rangle-(\langle\hat{p}_1\rangle-\langle\hat{p}_2\rangle)\langle\hat{q}_d\rangle-(\langle\hat{p}_1\rangle+\langle\hat{p}_2\rangle)\langle\hat{q}_c\rangle)&\nonumber\\&
+\frac{1}{2}(\langle\hat{q}_1\otimes\hat{p}_2\rangle-\langle\hat{p}_1\otimes\hat{q}_2\rangle)+\frac{1}{2}(\langle\hat{q}_d\rangle\langle\hat{p}_c\rangle-\langle\hat{q}_c\rangle\langle\hat{p}_d \rangle).&
\end{eqnarray}

The expectation values of the quadrature observables and square of the quadrature observables for reference modes can be computed directly (see Appendix~\ref{reference} for details). For the single mode states of the two-mode Gaussian state we have already discussed the method to estimate $\langle \hat{q}_k\rangle,$ $\langle \hat{p}_k\rangle,$ $\langle \hat{q}_k^2\rangle,$ $\langle \hat{p}_k^2\rangle,$ $\langle \hat{q}_k\hat{p}_k+\hat{p}_k\hat{q}_k\rangle.$ Using commutation relation $[q_k ,p_l]=\imath\delta_{kl}\openone,$ and for the unbiased reference state $\langle\hat{q}_r\hat{p}_r\rangle=\imath/2,$ and $\langle\hat{p}_r\hat{q}_r\rangle=-\imath/2,$ we can simplify Eqs.~$\eqref{stokes6}$, and $\eqref{stokes7}.$ As one can estimate $\langle\hat{S}_1^{2}(\phi_1=0)\rangle$ and $\langle(\hat{S}_1^{2}(\phi_2=0)\rangle$ from the experimental data, the two unknown quantities $\langle\hat{q}_1\otimes\hat{q}_2\rangle$ and $\langle\hat{p}_1\otimes\hat{p}_2\rangle$ can be computed by solving the two linear equations~$\eqref{stokes6}$, and $\eqref{stokes7}.$ Similarly, for biased reference state, one can obtain $\langle\hat{q}_r\hat{p}_r\rangle=\imath/2,$ and $\langle\hat{p}_r\hat{q}_r\rangle=-\imath/2$ for certain choices of parameters (for details, see the explanation  after Eq.~$\eqref{append6}$).

Similarly, following such a method one can estimate $\langle\hat{q}_1\otimes\hat{p}_2\rangle$ and $\langle\hat{p}_1\otimes\hat{q}_2\rangle$ by solving the two linear equations~$\eqref{stokes8}$, and $\eqref{stokes10},$ for certain choices of the pair of reference states. 
	The expectation values  $\langle\hat{q}_1\otimes\hat{p}_2\rangle$ and
	$\langle\hat{p}_1\otimes\hat{q}_2\rangle$ can be obtained by computing
	expectation values of  $\hat{S}_1^{2}(\phi_2=\frac{\pi}{4})$ and 
	$\hat{S}_1(\phi_1=0)\otimes \hat{S}_1(\phi_2=0),$ along with the
	previously estimated expectation values of the set of quadrature
	observables. In the case of reference states for which the first moments are zero, i.e., $\langle\hat{q}_{c/d}\rangle=0$ and/or $\langle\hat{p}_{c/d}\rangle=0,$ the same quantities can be estimated after solving Eqs.~$\eqref{stokes8}$, and $\eqref{stokes9}.$ In such a case, additionally, the expectation value of $\hat{S}_3(\phi_1=0, \phi_2=0)$ must be obtained from the experimental data.

Thus, for a given reference state, the elements of the covariance matrix
	associated to each of the single modes of the two-mode Gaussian state can
	be estimated for a single measurement setup with three phase shifts. In
	addition, considering the measurement setup in Figure~\ref{figg1}(b), for
	$\phi_1=0$ and $\phi_2=0, \frac{\pi}{4},$ the elements of the matrix $C$ are
	computed. 
	Once we know the elements of the covariance
	matrix, we can test \eqref{phys31aa} and quantify entanglement in the
	two-mode Gaussian state by computing $\text{det} A,$  $\text{det} B,$
	$\text{det} C,$ and $\text{det}\Gamma_{\rho_{12}}.$
Note that, the choice of phase shifts associated to Stokes-like measurements is not unique. It depends on the parameters of the reference state. 
However, as we don't require any apriori knowledge of the first moments and the second moments of the unknown two-mode Gaussian state, the scheme discussed here is universal. On the other hand, following this scheme, to test the separability criterion~$\eqref{phys31aa}$  one needs to estimate all elements of the covariance matrix as well as the first moments of the two-mode Gaussian state. Thus, the proposed scheme leads to the full state tomography. This may be due to the fact that  the measurements are performed on a single copy of the two-mode Gaussian state at a time. Similar results were obtained in the case of finite-dimensional systems. According to the Refs.~\cite{Carmeli16, Lu16}
 (for finite-dimensional systems)
, any universal entanglement detection scheme for a given bi-partite system, which
	uses single copy of the state at a time amounts to the full state tomography. However, for continuous variable systems, e.g., two-mode Gaussian states, we are not aware of such a result. 
	In the next section (Sec.~\ref{sec4}), we provide a scheme to estimate 
	$\det A, \det B, \det C,$ and $\det {\Gamma}_{{\rho}_{12}}$ without performing the  full state tomography.
	However, measurements on two copies of the
	two-mode Gaussian state are necessary. It is evident from the above discussion (related to Figure \ref{figg1}) that the aforesaid method of universal entanglement detection in two-mode Gaussian states is achieved here via LOCC only. On the other hand, measurement in an entangled basis is used in Sec. \ref{sec4} while dealing with two copies of the two-mode Gaussian states.

 Note that, in case of standard homodyne detection (for example in Refs.~\cite{D'Auria05, Mihaescu_2020}) the intensity of signal state
	must be less than the intensity of the reference state. Thus, it is may
	not be a feasible task to look for a reference state which has a higher
	intensity than the intensity of a signal state of undefined photon
	numbers, e.g., macroscopic states~\cite{Ruppert16}. In such a case,
	standard homodyne detection is difficult to perform. In Ref.~\cite{Ruppert16}, the
	elements of covariance matrix of a single mode macroscopic Gaussian state
	are estimated by performing Stokes-like measurements with a low-intensity
	reference state. In such a scheme, a strong coherent reference state can
	be replaced with a low-intensity displaced squeezed thermal reference
	state. In short, Stokes-like measurements generalize homodyne detection.
	The similar method we follow here to estimate elements of the covariance
	matrix of a two-mode Gaussian state. 

 \begin{figure}[ht!]
  \includegraphics[trim=250 275 250 300 width=1.75\linewidth]{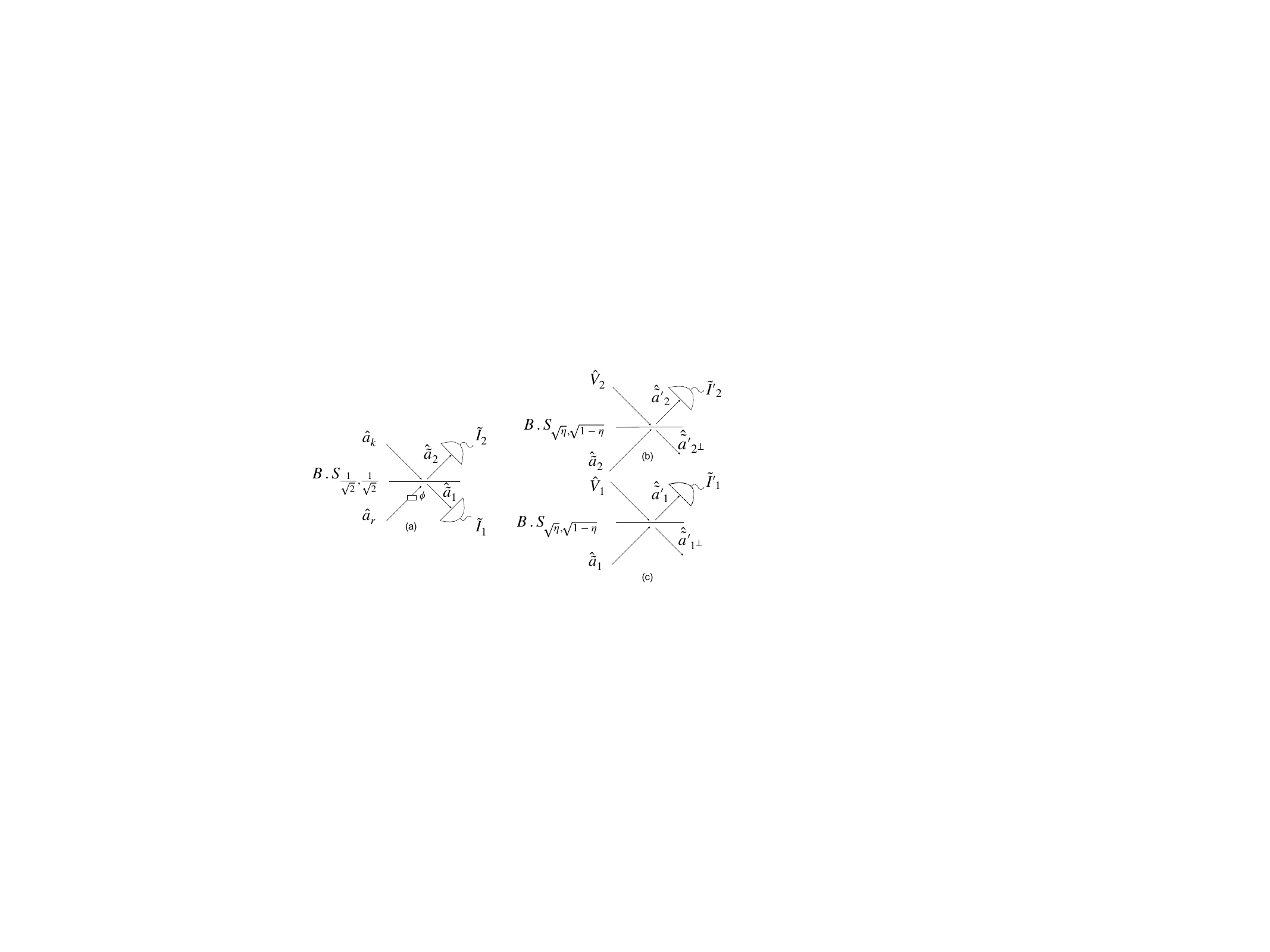}
\caption{
	}
\label{figg10}
\end{figure}

 \section{Robustness of our scheme}
 \label{sec4}
 In the preceding discussion, we have assumed ideal, perfectly efficient detectors. However, in a realistic experimental scenario, this assumption does not hold. Photon losses at the detectors are inevitable, and as a consequence, the measured statistics deviate from the ideal case and cannot be considered fully faithful. Therefore, when applying our scheme for detecting entanglement in an arbitrary two-mode Gaussian state, it is essential to account for such experimental imperfections.

In this work, we consider a specific model of detection inefficiency. An imperfect detector can be modeled by inserting a beam splitter just in front of the detector. In such a case, one of the output ports of this beam splitter is not detected, while the transmitted port is directed toward the detector. More precisely, for a  beam splitter characterized by transmittivity  $\sqrt{\eta}$ and reflectivity $\sqrt{1-\eta}$ (see Fig.~\ref{figg10}), we write 
 $\hat{\tilde{a}}_j'=\sqrt{\eta}\hat{\tilde{a}}_j+\sqrt{1-\eta}\hat{V}_j,$  where $j=1, 2,$ and $\hat{\tilde{a}}_j'$ is the annihilation operator associated to the detected output and $\hat{V}_j$ is annihilation operator of the vacuum state. Note that, the reflected output with annihilation operator $\hat{\tilde{a}}_{j^{\perp}}'=-\sqrt{1-\eta}\hat{\tilde{a}}_j+\sqrt{\eta}\hat{V}_j,$ where $j^{\perp}$ is orthogonal to j-th mode, is considered as not detected.Notably, such a beam splitter transformation preserves the Gaussian nature of the state; further details are provided in Appendix~\ref{robustness}.

\noindent 
Although our scheme is applicable to any two-mode Gaussian state, we illustrate its robustness using a symmetric two-mode squeezed vacuum (TMSV) input with squeezing parameter $r$. We compute the critical detection efficiency, denoted by  \(\eta_{\mathrm{critical}}\), below which the separability criterion~$\eqref{phys31aa}$ for TMSV state can no longer be violated.  These values are calculated for various squeezing parameters $r$
 and are summarized in Table~\ref{table:critical_eta}.
\begin{table}[h!]
\centering
\renewcommand{\arraystretch}{1.2}
\begin{tabular}{c|c}
\hline
\textbf{Squeezing parameter~($r$)}~ &~ \textbf{Critical detection efficiency ($\eta_{\text{critical}}$)} \\ \hline
0.5  & 0.848 \\
1.0  & 0.795 \\
1.5  & 0.704 \\
2.0  & 0.576 \\
2.5  & 0.433 \\ \hline
\end{tabular}
\caption{Critical detection efficiency $\eta_{\text{critical}}$ as a function of the squeezing parameter $r$.}
\label{table:critical_eta}
\end{table}

\section{Estimation of determinants without the full state tomography} \label{sec5}
\par In this section we discuss a different scheme of testing separability criterion \eqref{phys31aa}, without doing a tomography of the state. Thus, the certification of entanglement can be recast as a task to compute determinants of block matrices $A$, $B,$ $C,$ and the covariance matrix $\Gamma_{\rho_{12}}$  with less resources than the resources required for the full state tomography. 

\par From the definition
of the Wigner function $\eqref{Gauss}$ of the two-mode Gaussian state one can
check that 
 \begin{equation}
 \label{phys31a}
 (4\pi)^2\int_{-\infty}^{+\infty}\int_{-\infty}^{+\infty}\int_{-\infty}^{+\infty}\int_{-\infty}^{+\infty}\dd
	 p_2 \dd q_2 \dd p_1 \dd q_1\; W^2_{\rho_{12}}
	 (\hat{q}_1, \hat{p}_1, \hat{q}_2,
	 \hat{p}_2)=\frac{1}{\sqrt{\det\Gamma_{\rho_{12}}}}
 \end{equation} 
 Interestingly, the left-hand side of the equality is defined as
 $\text{Tr}\rho_{12}^2.$ Similarly, we have  $\text{Tr}\rho_{1}^2=1/\sqrt{\det
 A},$ and $\text{Tr}\rho_{2}^2=1/\sqrt{\det B},$ where $\rho_{1}$ and
 $\rho_{2}$ can be obtained by tracing out mode 2 and mode 1 of the
 two-mode Gaussian state $\rho_{12}$, respectively. Therefore, experimentally realizable schemes to estimate $\text{Tr}\rho_{1}^2, \text{Tr}\rho_{2}^2,$ and $\text{Tr}\rho_{12}^2$ will yield $\det A, \det B,$ and $ \det\Gamma_{\rho_{12}},$ respectively.

To this end we consider the SWAP operator $\hat{\mathbb{S}}$ :
$\hat{\mathbb{S}}(|m\rangle_1 \otimes |n\rangle_2) = |n\rangle_1 \otimes |m\rangle_2,$ where $|k\rangle_j$ is the k-th Fock state of mode $j$. 
It can now be checked that: $\tr [\hat{\mathbb{S}}(\rho_1\otimes \rho_1)] = \tr \rho_1^2$,  $\tr [\hat{\mathbb{S}}(\rho_2 \otimes \rho_2)] = \tr \rho_2^2$. 
Such a SWAP operator  acting on the two-mode system together  is known to be self-adjoint, although  physical realization of measurement of the SWAP operator on the two-mode system may turn out to be quite difficult (see Ref.~\cite{Wang_2001}). Similar argument may also be provided for the physical realization of measurement of the $(2 + 2)$-modes SWAP operator $\hat{\mathbb{S}}^{\prime}.$ Thus  $\tr {\rho}_{12}^2$ can be obtained from  the relation $\tr [\hat{\mathbb{S}}^{\prime}({\rho}_{12} \otimes {\rho}_{12})] = \tr {\rho}_{12}^2$, where $\hat{\mathbb{S}}^{\prime}$ is the SWAP operator acting on $2 + 2$ modes {\it i.e.}, $\hat{\mathbb{S}'}((|n\rangle_1 \otimes |m\rangle_1') \otimes (|r\rangle_2 \otimes |s\rangle_2')) = (|r\rangle_1 \otimes |s\rangle_1') \otimes (|n\rangle_2 \otimes |m\rangle_2')$ for all $n, m, r, s = 0, 1, \ldots .$
As $|n\rangle_1 \otimes |n\rangle_2$ (for $n = 0, 1, \ldots$) and $(1/{\sqrt{2}})(|n\rangle_1 \otimes |n + k\rangle_2 + |n + k\rangle_1 \otimes |n\rangle_2)$ (for $n = 0, 1, \ldots$ and $k = 1, 2, \ldots$) are the eigen states of the SWAP operator corresponding to the eigenvalue $+ 1$ while $(1/{\sqrt{2}})(|n\rangle_1 \otimes |n + k\rangle_2 - |n + k\rangle_1 \otimes |n\rangle_2)$ (for $n = 0, 1, \ldots$ and $k = 1, 2, \ldots$) are eigenstates of the operator corresponding to the eigenvalue $- 1$, therefore, measurement of the SWAP operator $\hat{\mathbb{S}}$ would correspond to the projective measurement in the basis $\{|n\rangle_1 \otimes |n\rangle_2 : n = 0, 1, \ldots\} \bigcup \{(1/{\sqrt{2}})(|n\rangle_1 \otimes |n + k\rangle_2 + |n + k\rangle_1 \otimes |n\rangle_2 : n = 0, 1, \ldots;  k = 1, 2, \ldots\} \bigcup  \{(1/{\sqrt{2}})(|n\rangle_1 \otimes |n + k\rangle_2 - |n + k\rangle_1 \otimes |n\rangle_2 : n = 0, 1, \ldots;  k = 1, 2, \ldots\}$. On the other hand, measurement of the SWAP operator $\hat{\mathbb{S}}^{\prime}$ would amount to measurement in the basis $\{(|n\rangle_1 \otimes |m\rangle_{1'}) \otimes (|n\rangle_2 \otimes |m\rangle_{2'}) : n, m = 0, 1, \ldots\} \bigcup \{(1/{\sqrt{2}})((|n\rangle_1 \otimes |m\rangle_{1'}) \otimes (|n + k\rangle_2 \otimes |m + l\rangle_{2'}) + (|n + k\rangle_1 \otimes |m + l\rangle_{1'}) \otimes (|n\rangle_2 \otimes |m\rangle_{2'})) : n, m = 0, 1, \ldots; k, l = 1, 2, \ldots\} \bigcup  \{(1/{\sqrt{2}})((|n\rangle_1 \otimes |m\rangle_{1'}) \otimes (|n + k\rangle_2 \otimes |m + l\rangle_{2'}) - (|n + k\rangle_1 \otimes |m + l\rangle_{1'}) \otimes (|n\rangle_2 \otimes |m\rangle_{2'})) : n, m = 0, 1, \ldots; k, l = 1, 2, \ldots\}$ -- a global measurement on all the four modes $1, 1', 2, 2'$ together.  Note that such a measurement is practically impossible to perform as it may seem to require photon number resolving detectors with infinite resolution, although recent experimental work \cite{Nguyen21} does provide implementation of $(1 + 1)$-mode SWAP operator as a unitary operator on two motional states in  a system of trapped $^{171}Yb^+$ ions.  In case there is some restriction on the average photon numbers of the input two-mode Gaussian states, one can, in principle, perform the aforesaid projective measurement with restricted photon number resolving detectors. 
We see that the measurement of $\hat{\mathbb{S}}$ on the  two copies of the single-mode reduced density matrix $\rho_1$, two copies  of the single-mode reduced density matrix $\rho_2$, and also measurement of $\hat{\mathbb{S}}^{\prime}$ on two copies of the state $\rho_{12}$ provide us the values of the three local symplectic invariants $\det A$, $\det B$, and $\det {\Gamma_{\rho_{12}}}$. Thus, we are left with the computation of $\det C.$ Three different methods of obtaining elements of  matrix $C$ are given below.

$\it{Method ~1:}$  On single copy of the unknown two-mode Gaussian state Alice (in 
possession of mode 1) performs at random measurements of one of 
the two observables $\hat{q}_1$, $\hat{p}_1$. Also, on that same copy 
of the two-mode Gaussian state, Bob (in possession of mode 2) performs 
at random measurements of one of the two observables 
$\hat{q}_2$, $\hat{p}_2$. They will then communicate classically 
regarding the choice of their measurements. By this method, Alice and 
Bob together can find out the matrix C, and thereby another local 
symplectic invariant, $\det C$.

$\it{Method ~2:}$ If we assume that the covariance matrix is of Simon type; i.e., $A=
\lambda I_2,\, B = \mu I_2,$ and $C= \begin{bmatrix} s &0 \\ 0& t \end{bmatrix}$,
	then the computation can be simplified. ($I_2$ denotes the $2\times2$ identity matrix). Note that $\det A,\, \det B$, and
	$\det C$ are invariant under local symplectic transformations. Hence by the method described above, we may calculate $\lambda$ and $\mu$ which are positive
	square roots of $\det A$ and $\det B$ respectively.  To calculate $|\det C|$ we use the fact that  due to the special structure of Simon's form $\det\Gamma_{\rho_{12}} = \det(\lambda\mu I_2 - CC^t)$. Simple algebra shows that  $\det\Gamma_{\rho_{12}} = (\lambda\mu)^2 - \lambda\mu(s+t) +st$. Similarly we may apply a Gaussian rotation matrix of the form $\begin{bmatrix} \cos \theta I_2 & \sin\theta I_2 \\-\sin\theta I_2 & \cos\theta I_2 
	\end{bmatrix}.$ For the value $\theta = \frac{\pi}{4}$ we observe that the marginal covariance matrix with respect to mode 1 takes the form $(\lambda+\mu)I_2  - (C+C^t)$.  Determinant of this can be calculated by the previous method which will take the form $\det((\lambda+\mu)I_2  - (C+C^t))= (\lambda+\mu)^2 - 2(\lambda+\mu)(s+t) +4st$. Since the value of $\lambda$ and $\mu$ are known, and the determinants of the left hand sides can be estimated, $s$ and $t$ can be calculated by solving the two equations. As a result the $\det C$ can also be calculated.

  \begin{figure}
  \centering
  \includegraphics[trim=200 275 400 300 width=1.75\linewidth]{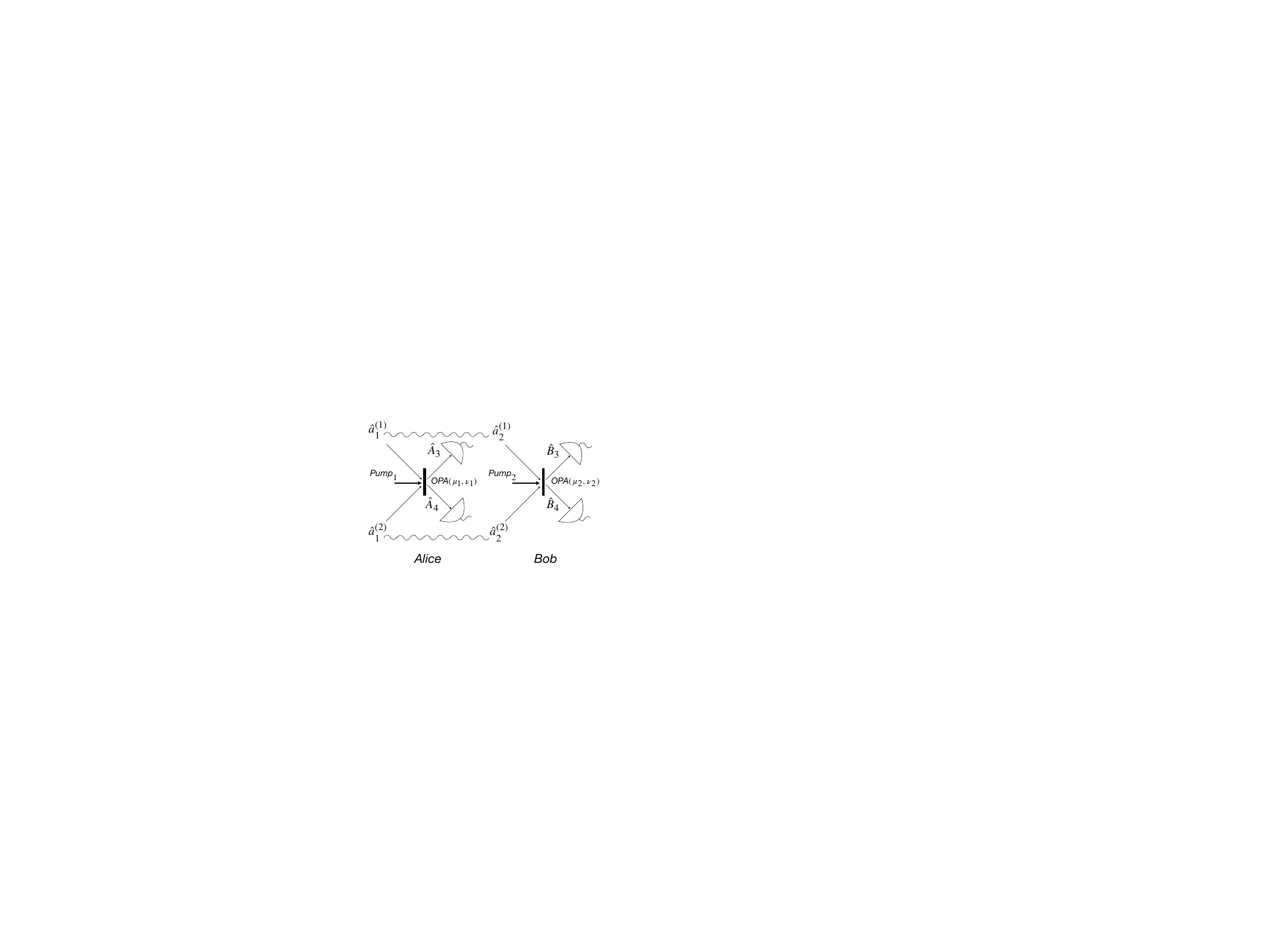}
  \vspace{-1 cm}
\caption{A schematic diagram of the measurement setup to estimate all elements of  the matrix $C$ associated to a two-mode Gaussian state. Two copies of the state are shared between Alice and Bob. $(\hat{a}^{(1)}_{1}, \hat{a}^{(2)}_{1})$ and $(\hat{a}^{(1)}_{2}, \hat{a}^{(2)}_{2})$ 
are two pairs of annihilation operators associated to the two pairs of input modes of the OPAs on Alice's and Bob's side, respectively. Note that, the correlated modes are $\hat{a}^{(j)}_{1},$ and $\hat{a}^{(j)}_{2}.$ $\text{Pump}_1,$ and $\text{Pump}_2$ are sources of pump beams incident on the OPA on Alice's and Bob's side, respectively.  The output modes of the OPA on Alice's and Bob's side are represented by annihilation operators $\hat{A}_k,$ and $\hat{B}_k,$ respectively. Here $j\in\{1, 2\},$ and $k\in\{3, 4\}.$}
\label{figg2}
\end{figure}

$\it{Method ~3:}$ Here, our aim is to estimate the elements of the matrix $C$ by performing Stokes-like measurements. To implement  such a scheme, Alice and Bob are given two copies of a two-mode Gaussian state and each of them possesses  one OPA~\cite{Plick-NJP-2010}. In addition, we assume that the first moments of the  Gaussian state are zero, i.e., $\langle \hat{q}_i\rangle=\langle\hat{p}_i\rangle=0$ for $i = 1, 2.$ This, one can take without loss of generality, as, otherwise by phase-space displacement one can bring the two-mode Gaussian state in that form. The amount of the aforesaid phase-space displacements can be obtained by first finding out the mean values ${\langle}\hat{q}_1{\rangle}$, ${\langle}\hat{p}_1{\rangle}$, ${\langle}\hat{q}_2{\rangle}$, ${\langle}\hat{p}_2{\rangle}$ (following, as for example, the method described in Figure \ref{figg1}(a)), and thereby, applying the corresponding phase-space displacements to bring down the aforesaid mean values to zero. 

The two modes of the j-th copy of the Gaussian state on Alice's and Bob's side are represented by  the annihilation operators $\hat{a}^{(j)}_{1},$ and $\hat{a}^{(j)}_{2},$ respectively (see Figure~\ref{figg2}). Note that, the correlated modes are denoted by the annihilation operators $(\hat{a}^{(j)}_{1}, \hat{a}^{(j)}_{2})$. Input modes of the each of the two OPAs are represented by  the annihilation operators $(\hat{a}^{(1)}_{i}, \hat{a}^{(2)}_{i}).$ After the interference separately at the two OPAs, the annihilation operators of the outputs on Alice's and Bob's side are $\hat{A}_k$ and $\hat{B}_k,$ respectively. If $\mu_l=\cosh g_l,$ and $\nu_l=e^{\imath\Phi_l}\sinh g_l,$ where $g_l$ and $\Phi_l$ are parametrical strength, which depends on the intensity of the pump beam as well as the  nonlinearity of the OPA crystal and  phase of the pump beam, respectively, then by definition, (see Refs.~\cite{Plick-NJP-2010, Ma18})  for $l\in\{1, 2\},$ we can write
\begin{eqnarray}
 \label{append8}
&\hat{A}_3=\mu_1\hat{a}^{(2)}_1+\nu_1\hat{a}^{(1)\dagger}_1~~ \hat{A}_4=\mu_1\hat{a}^{(1)}_1+\nu_1\hat{a}^{(2)\dagger}_1&\nonumber\\
&\hat{B}_3=\mu_2\hat{a}^{(2)}_2+\nu_2\hat{a}^{(1)\dagger}_2~~ \hat{B}_4=\mu_2\hat{a}^{(1)}_2+\nu_2\hat{a}^{(2)\dagger}_2,&
 \end{eqnarray}
where, on Alice's side and Bob's side, the OPA transformations are parametrized by  $(\mu_1=\cosh g_1, \nu_1=e^{\imath\Phi_1}\sinh g_1),$ and $(\mu_2=\cosh g_2, \nu_2=e^{\imath\Phi_2}\sinh g_2),$ respectively. 

To compute the elements of the matrix $C,$ we need to estimate expectation values of a set of Stokes-like measurements, e.g., $\imath(\hat{A}_3^\dagger\hat{B}_3-\hat{B}_3^\dagger\hat{A}_3),$ $\hat{A}_3^\dagger\hat{B}_3+\hat{B}_3^\dagger\hat{A}_3,$ $\imath(\hat{A}_3^\dagger\hat{B}_4-\hat{B}_4^\dagger\hat{A}_3),$ and $\hat{A}_3^\dagger\hat{B}_4+\hat{B}_4^\dagger\hat{A}_3.$ Using $\eqref{append8}$
 and replacing the annihilation and the creation operators with quadrature observables, a pair of expectation values, which requires measurements jointly on two different modes, is given as follows:
 \begin{eqnarray}
 \label{append9}
&\langle \imath(\hat{A}_3^\dagger\hat{B}_3-\hat{B}_3^\dagger\hat{A}_3)\rangle=M_1(\langle \hat{q}_1\otimes \hat{q}_2\rangle+\langle \hat{p}_1\otimes \hat{p}_2\rangle)+N_1(\langle \hat{q}_1\otimes \hat{p}_2\rangle-\langle \hat{p_1}\otimes \hat{q}_2\rangle),&
 \end{eqnarray}
  \begin{eqnarray}
   \label{append10}
&\langle (\hat{A}_3^\dagger\hat{B}_3+\hat{B}_3^\dagger\hat{A}_3)\rangle=M_2(\langle \hat{q}_1\otimes \hat{q}_2\rangle+\langle \hat{p}_1\otimes \hat{p}_2\rangle)+N_2(\langle \hat{q}_1\otimes \hat{p}_2\rangle-\langle \hat{p_1}\otimes \hat{q}_2\rangle),&
 \end{eqnarray}
where $M_1,  N_1, M_2,$ and $N_2$ \footnote{$M_1=\sinh g_1\sinh g_2\sin(\Phi_1-\Phi_2),$  
$N_1=-\cosh g_1\cosh g_2+\sinh g_1\sinh g_2\cos(\Phi_1-\Phi_2),$ 
$M_2=\cosh g_1\cosh g_2+\sinh g_1\sinh g_2\cos(\Phi_1-\Phi_2),$ 
$N_2=-\sinh g_1\sinh g_2\sin(\Phi_1-\Phi_2)$ } are constants. For another pair of Stokes-like measurements we obtain
\begin{eqnarray}
 \label{append11}
&\langle \imath(\hat{A}_3^\dagger\hat{B}_4-\hat{B}_4^\dagger\hat{A}_3)\rangle=M_1'(\langle \hat{q}_1\otimes \hat{q}_2\rangle-\langle \hat{p}_1\otimes \hat{p}_2\rangle)+N_1'(\langle \hat{q}_1\otimes \hat{p}_2\rangle+\langle \hat{p_1}\otimes \hat{q}_2\rangle),&
 \end{eqnarray}
\begin{eqnarray}
 \label{append12}
&\langle (\hat{A}_3^\dagger\hat{B}_4+\hat{B}_4^\dagger\hat{A}_3)\rangle=M_2'(\langle \hat{q}_1\otimes \hat{q}_2\rangle-\langle \hat{p}_1\otimes \hat{p}_2\rangle)+N_2'(\langle \hat{q}_1\otimes \hat{p}_2\rangle+\langle \hat{p_1}\otimes \hat{q}_2\rangle),&
 \end{eqnarray}
where $M_1',  N_1', M_2',$ and $N_2'$ \footnote{$M_1'=-\cosh g_1\sinh g_2\sin\Phi_2+\sinh g_1\cosh g_2\sin \Phi_1$,
$N_1'=\cosh g_1\sinh g_2\cos\Phi_2-\sinh g_1\cosh g_2\cos\Phi_1,$  
$M_2'=\cosh g_1\sinh g_2\cos\Phi_2+\sinh g_1\cosh g_2\cos\Phi_1,$ 
$N_2'=\sinh g_1\cosh g_2\sin\Phi_1+\cosh g_1\sinh g_2\sin \Phi_2$}  are constants. Here we consider the fact that  $\langle \hat{q}^{(j)}_i\rangle=\langle\hat{p}^{(j)}_i\rangle=0.$ Also,  $\langle \hat{q}^{(j)}_1\otimes \hat{q}^{(j)}_2\pm \hat{p}^{(j)}_1\otimes \hat{p}^{(j)}_2\rangle=\langle \hat{q}_1\otimes \hat{q}_2\pm \hat{p}_1\otimes \hat{p}_2\rangle=\langle \hat{q}_1\otimes \hat{q}_2\rangle\pm\langle \hat{p}_1\otimes \hat{p}_2\rangle.$ Similarly, we use $\langle \hat{q}^{(j)}_1\otimes \hat{p}^{(j)}_2\mp \hat{p}^{(j)}_1\otimes \hat{q}^{(j)}_2\rangle=\langle \hat{q}_1\otimes \hat{p}_2\mp \hat{p}_1\otimes \hat{q}_2\rangle=\langle \hat{q}_1\otimes \hat{p}_2\rangle\mp\langle \hat{p}_1\otimes \hat{q}_2\rangle.$ This is so because $\langle \hat{q}^{(j)}_1\otimes \hat{q}^{(j)}_2\rangle,$  $\langle \hat{q}^{(j)}_1\otimes \hat{p}^{(j)}_2\rangle,$ $\langle \hat{q}^{(j)}_2\otimes \hat{p}^{(j)}_1\rangle,$ and $\langle \hat{p}^{(j)}_1\otimes \hat{p}^{(j)}_2\rangle$ are independent, where $j\in\{1, 2\}.$ Note that, $\langle \hat{q}^{(s)}_m\otimes \hat{p}^{(t)}_n\rangle=\langle \hat{q}^{(s)}_m\rangle\langle \hat{p}^{(t)}_n\rangle=0,$ where $s\neq t.$ As expectation values of the Stokes-like measurements can be estimated from experiments, all the elements of matrix $C$ can be obtained by solving Eqs.~$\eqref{append9}, \eqref{append10}, \eqref{append11},$ and $\eqref{append12}.$ It is worth mentioning that the elements of the $C$ matrix can also be found by computing expectation values of generators of SU(1, 1) group~\cite{Caves_2020}.


\par Note that, the total number of measurements to obtain $\det A, \det B,$ and $ \det\Gamma_{\rho_{12}}$ by SWAP operations is three. In addition, the required numbers of measurements for computing $\det C$ are  four (random measurements of quadrature observables), one (for special forms of the matrices $A, B, C$, one can perform SWAP operation $\mathbb{S}'$), and four (for given first moments of the quadratures of the state, four Stokes-like measurements are required here) for Method 1, Method 2, and Method 3, respectively. Although, for Methods 1, and 3, the $\det C$ can be computed from the elements of matrix $C,$ but  $\det A,$ $\det B,$ and $\det \Gamma_{\rho_{12}}$ are found without obtaining the elements of the matrices. Precisely, we don't need to estimate $\langle \hat{q}^{2}_m\rangle,$ $\langle \hat{p}^{2}_m\rangle,$ and $\frac{1}{2}\langle \hat{q}_m\hat{p}_m+\hat{p}_m\hat{q}_m\rangle .$ We have thereby, described a scheme to test the separability criterion $\eqref{phys31aa}$
for two-mode Gaussian state without the full state tomography. Thus, resource-wise, the scheme is more economical than the full state tomography.

\section{Discussion}\label{s5}
\par In this paper we present altogether four schemes for detection of entanglement in an unknown
two-mode Gaussian state. The first three schemes are described with measurements on a single copy of the state at a time. An identical set of {\em five} measurements is required to execute the schemes demonstrated in Section~\ref{s2} and in Appendix~\ref{app2}. However, it will be interesting to find a set of experimentally measurable quantities for such {\em five}  measurements which will be resource-wise more economical than a set of measurements for the full state `tomography' although single copy usage of the sate does not seem to be better than state tomography in terms of resource requirements. In Section~\ref{s3}, we provide an experimental friendly scheme to test the separability criterion~\eqref{phys31aa}. The elements of the covariance matrix of the unknown Gaussian state are estimated by measuring intensity at the outputs of the interferometric setups (see Figure~\ref{figg1}). We conjecture that as in such cases the manipulations are made on single copy of the state, the scheme leads to the full state tomography~\cite{Carmeli16}. In this work (see Section~\ref{sec4}), we evaluate the robustness of our scheme by accounting for detector inefficiency. For a specific choice of reference states in our homodyne detection setup, we observe that as the squeezing parameter of the symmetric two-mode squeezed vacuum (TMSV) state increases, the critical detection efficiency \(\eta_{\text{critical}}\) required to violate the separability criterion decreases. This implies that for a given 
$r$, entanglement is successfully detected provided that the experimental detection efficiency is greater than \(\eta_{\text{critical}}\).
Finally, in Section~\ref{sec5}, we discuss a scheme by considering measurements on two copies of the state at a time. Interestingly, one can exploit the structure of SWAP operator and estimate the determinant (taking two copies of the state at a time) of matrices $A, B.$ In addition, considering OPA transformations on two copies of the two-mode Gaussian state, we compute $\det C.$ As it does not require knowledge of each and every parameter of the state, the scheme does not amount to the full state tomography.  It is worth mentioning that  the schemes described in present work, also give rise to the estimation of the measure of entanglement in the two-mode Gaussian state. For example, one can quantify the entanglement of a two-mode Gaussian state in terms of the logarithmic negativity (see the discussion below Eq.~\eqref{phys31aa}). Note that, the advantage of the scheme presented in Section~\ref{s3} over other schemes with measurements on two copies of the state at a time lies in the fact that the former can be more robust against errors than the later if the source which prepares the unknown two-mode Gaussian state is  erroneous. In addition, in Ref.~\cite{Ruppert16}, it was shown that the robustness of the estimation of covariance matrix of the single mode Gaussian state does not depend on the choice of biased or unbiased displaced squeezed thermal state (reference state) for about $10^4-10^5$ measurements. We expect the same for the two-mode Gaussian state. However, we need to investigate the robustness of the scheme by considering imperfect detectors, and mode-matching error between signal states and reference states interfering on beam-splitters. We also need to look for realistic measurements rather than the measurement of the $(1 + 1)$-modes as well as $(2 + 2)$-modes SWAP operators $\hat{\mathbb{S}},$ and $\hat{\mathbb{S}}^{\prime},$ respectively while dealing with two copies of the two-mode Gaussian states together.

In future, we would like to extend the scheme in Section~\ref{s3} for multimode systems together with finding out the optimal universal entanglement witnessing scheme for any given multi-mode Gaussian states. It is worth mentioning that recently, a measurement-device-independent scheme was proposed by Abiuso {\it et al.}~\cite{Abiuso2021} based on the witness provided in Ref.~\cite{Duan00}. However, the scheme is not universal, as partial information regarding the two-mode Gaussian state is needed to transform it into a "standard form" for testing. Since our entanglement detection scheme does not assume a specific structure for the two-mode Gaussian state, a measurement-device-independent method based on Stokes-like operators will be presented in a forthcoming paper.

\section*{Funding}
 J.K. is supported in part by KIAS Advanced Research Program
(No. CG014604). R.S. acknowledges financial supports from {\sf SERB MATRICS
MTR/2017/000431} and {\sf
DST/ICPS/QuST/Theme-2/2019/General} Project number {\sf
Q-90}.
\section*{Acknowledgments}
The authors would like to thank Paulina Marian for pointing out a relevant reference~\cite{Marian18}.
\section*{Disclosures} The authors declare no conflicts of interest.
\bibliography{biblio}

@article{Caves_2020,
author = {Caves, Carlton M.},
title = {Reframing {$SU(1,1)$} Interferometry},
journal = {Advanced Quantum Technologies},
volume = {3},
number = {11},
pages = {1900138},
doi = {https://doi.org/10.1002/qute.201900138},
url = {https://onlinelibrary.wiley.com/doi/abs/10.1002/qute.201900138},
eprint = {https://onlinelibrary.wiley.com/doi/pdf/10.1002/qute.201900138},
year = {2020}
}

@article{Ma18,
author = {Xiaoping Ma and Chenglong You and Sushovit Adhikari and Elisha S. Matekole and Ryan T. Glasser and Hwang Lee and Jonathan P. Dowling},
journal = {Opt. Express},
number = {14},
pages = {18492--18504},
publisher = {OSA},
title = {Sub-shot-noise-limited phase estimation via SU(1,1) interferometer with thermal states},
volume = {26},
month = {Jul},
year = {2018},
url = {http://www.opticsexpress.org/abstract.cfm?URI=oe-26-14-18492},
doi = {10.1364/OE.26.018492},
}

@ARTICLE{Nguyen21,
       author = {{Nguyen}, Chi-Huan and {Tseng}, Ko-Wei and {Maslennikov}, Gleb and {Gan}, H.~C.~J. and {Matsukevich}, Dzmitry},
        title = "{Experimental SWAP test of infinite dimensional quantum states}",
      journal = {arXiv e-prints},
         year = 2021,
        month = mar,
          eid = {arXiv:2103.10219},
        pages = {arXiv:2103.10219},
archivePrefix = {arXiv},
       eprint = {2103.10219},
 primaryClass = {quant-ph},
       adsurl = {https://ui.adsabs.harvard.edu/abs/2021arXiv210310219N}
}

@article{Marian18,
	doi = {10.1088/1751-8121/aa9fae},
	url = {https://doi.org/10.1088/1751-8121/aa9fae},
	year = 2018,
	month = {jan},
	publisher = {{IOP} Publishing},
	volume = {51},
	number = {6},
	pages = {065301},
	author = {Paulina Marian and Tudor A Marian},
	title = {Einstein{\textendash}Podolsky{\textendash}Rosen-like separability indicators for two-mode Gaussian states},
	journal = {Journal of Physics A: Mathematical and Theoretical},
}

@article{Duan00,
  title = {Inseparability Criterion for Continuous Variable Systems},
  author = {Duan, Lu-Ming and Giedke, G. and Cirac, J. I. and Zoller, P.},
  journal = {Phys. Rev. Lett.},
  volume = {84},
  issue = {12},
  pages = {2722--2725},
  year = {2000},
  month = {Mar},
  doi = {10.1103/PhysRevLett.84.2722},
  url = {http://link.aps.org/doi/10.1103/PhysRevLett.84.2722},
  publisher = {American Physical Society}
}

@article{PhysRevA.65.052306,
  title = {Polarization squeezing and continuous-variable polarization entanglement},
  author = {Korolkova, Natalia and Leuchs, Gerd and Loudon, Rodney and Ralph, Timothy C. and Silberhorn, Christine},
  journal = {Phys. Rev. A},
  volume = {65},
  issue = {5},
  pages = {052306},
  numpages = {12},
  year = {2002},
  month = {Apr},
  publisher = {American Physical Society},
  doi = {10.1103/PhysRevA.65.052306},
  url = {https://link.aps.org/doi/10.1103/PhysRevA.65.052306}
}

@article{PhysRevA.67.012316,
  title = {Stokes-operator-squeezed continuous-variable polarization states},
  author = {Schnabel, Roman and Bowen, Warwick P. and Treps, Nicolas and Ralph, Timothy C. and Bachor, Hans-A. and Lam, Ping Koy},
  journal = {Phys. Rev. A},
  volume = {67},
  issue = {1},
  pages = {012316},
  numpages = {11},
  year = {2003},
  month = {Jan},
  publisher = {American Physical Society},
  doi = {10.1103/PhysRevA.67.012316},
  url = {https://link.aps.org/doi/10.1103/PhysRevA.67.012316}
}

@article{Plick-NJP-2010,
	doi = {10.1088/1367-2630/12/11/113025},
	url = {https://doi.org/10.1088/1367-2630/12/11/113025},
	year = 2010,
	month = {nov},
	publisher = {{IOP} Publishing},
	volume = {12},
	number = {11},
	pages = {113025},
	author = {William N Plick and Petr M Anisimov and Jonathan P Dowling and Hwang Lee and Girish S Agarwal},
	title = {Parity detection in quantum optical metrology without number-resolving detectors},
	journal = {New Journal of Physics}
}

@article{Carmeli16,
  title = {Verifying the Quantumness of Bipartite Correlations},
  author = {Carmeli, Claudio and Heinosaari, Teiko and Karlsson, Antti and Schultz, Jussi and Toigo, Alessandro},
  journal = {Phys. Rev. Lett.},
  volume = {116},
  issue = {23},
  pages = {230403},
  numpages = {5},
  year = {2016},
  month = {Jun},
  publisher = {American Physical Society},
  doi = {10.1103/PhysRevLett.116.230403},
  url = {https://link.aps.org/doi/10.1103/PhysRevLett.116.230403}
}

@article{Lu16,
  title = {Tomography is Necessary for Universal Entanglement Detection with Single-Copy Observables},
  author = {Lu, Dawei and Xin, Tao and Yu, Nengkun and Ji, Zhengfeng and Chen, Jianxin and Long, Guilu and Baugh, Jonathan and Peng, Xinhua and Zeng, Bei and Laflamme, Raymond},
  journal = {Phys. Rev. Lett.},
  volume = {116},
  issue = {23},
  pages = {230501},
  numpages = {5},
  year = {2016},
  month = {Jun},
  publisher = {American Physical Society},
  doi = {10.1103/PhysRevLett.116.230501},
  url = {https://link.aps.org/doi/10.1103/PhysRevLett.116.230501}
}

@article{Mihaescu_2020,
	doi = {10.1088/1367-2630/abd1ad},
	url = {https://doi.org/10.1088/1367-2630/abd1ad},
	year = 2020,
	month = {dec},
	publisher = {{IOP} Publishing},
	volume = {22},
	number = {12},
	pages = {123041},
	author = {Tatiana Mihaescu and Hermann Kampermann and Giulio Gianfelici and Aurelian Isar and Dagmar Bru{\ss}},
	title = {Detecting entanglement of unknown continuous variable states with random measurements},
	journal = {New Journal of Physics},
	}

@article{Wang_2001,
	doi = {10.1088/0305-4470/34/44/316},
	url = {https://doi.org/10.1088/0305-4470/34/44/316},
	year = 2001,
	month = {oct},
	publisher = {{IOP} Publishing},
	volume = {34},
	number = {44},
	pages = {9577--9584},
	author = {Xiaoguang Wang},
	title = {Continuous-variable and hybrid quantum gates},
	journal = {Journal of Physics A: Mathematical and General},
}

@Article{Peres96, 
  title = {Separability Criterion for Density Matrices},
  author = {Peres, Asher },
  journal = {Phys. Rev. Lett.},
  volume = {77},
  number = {8},
  pages = {1413--1415},
  numpages = {2},
  year = {1996},
  month = {Aug},
  doi = {10.1103/PhysRevLett.77.1413},
  publisher = {American Physical Society}
}

@article{Horodecki96,
    AUTHOR = {Horodecki, Micha{\l} and Horodecki, Pawe{\l} and Horodecki,
              Ryszard},
     TITLE = {Separability of mixed states: necessary and sufficient
              conditions},
   JOURNAL = {Phys. Lett. A},
  FJOURNAL = {Physics Letters. A},
    VOLUME = {223},
      YEAR = {1996},
    NUMBER = {1-2},
     PAGES = {1--8},
       DOI = {10.1016/S0375-9601(96)00706-2},
       URL = {http://dx.doi.org/10.1016/S0375-9601(96)00706-2},

}

@article{Simon00,
  title = {{P}eres-{H}orodecki Separability Criterion for Continuous Variable Systems},
  author = {Simon, R.},
  journal = {Phys. Rev. Lett.},
  volume = {84},
  issue = {12},
  pages = {2726--2729},
  year = {2000},
  month = {Mar},
  doi = {10.1103/PhysRevLett.84.2726},
  url = {http://link.aps.org/doi/10.1103/PhysRevLett.84.2726},
  publisher = {American Physical Society}
}

@article{cerf04,
  title = {How to Measure Squeezing and Entanglement of Gaussian States without Homodyning},
  author = {Fiur\'a\ifmmode \check{s}\else \v{s}\fi{}ek, Jarom\'{\i}r and Cerf, Nicolas J.},
  journal = {Phys. Rev. Lett.},
  volume = {93},
  issue = {6},
  pages = {063601},
  numpages = {4},
  year = {2004},
  month = {Aug},
  publisher = {American Physical Society},
  doi = {10.1103/PhysRevLett.93.063601},
  url = {https://link.aps.org/doi/10.1103/PhysRevLett.93.063601}
}

@article{vidal02,
  title = {Computable measure of entanglement},
  author = {Vidal, G. and Werner, R. F.},
  journal = {Phys. Rev. A},
  volume = {65},
  issue = {3},
  pages = {032314},
  numpages = {11},
  year = {2002},
  month = {Feb},
  publisher = {American Physical Society},
  doi = {10.1103/PhysRevA.65.032314},
  url = {https://link.aps.org/doi/10.1103/PhysRevA.65.032314}
}

@article{adesso08,
  title = {Genuine multipartite entanglement of symmetric Gaussian states: Strong monogamy, unitary localization, scaling behavior, and molecular sharing structure},
  author = {Adesso, Gerardo and Illuminati, Fabrizio},
  journal = {Phys. Rev. A},
  volume = {78},
  issue = {4},
  pages = {042310},
  numpages = {18},
  year = {2008},
  month = {Oct},
  publisher = {American Physical Society},
  doi = {10.1103/PhysRevA.78.042310},
  url = {https://link.aps.org/doi/10.1103/PhysRevA.78.042310}
}

@article{Giedke03,
  title = {Entanglement of Formation for Symmetric Gaussian States},
  author = {Giedke, G. and Wolf, M. M. and Kr\"uger, O. and Werner, R. F. and Cirac, J. I.},
  journal = {Phys. Rev. Lett.},
  volume = {91},
  issue = {10},
  pages = {107901},
  numpages = {4},
  year = {2003},
  month = {Sep},
  publisher = {American Physical Society},
  doi = {10.1103/PhysRevLett.91.107901},
  url = {https://link.aps.org/doi/10.1103/PhysRevLett.91.107901}
}

@article{Wolf04,
  title = {Gaussian entanglement of formation},
  author = {Wolf, M. M. and Giedke, G. and Kr\"uger, O. and Werner, R. F. and Cirac, J. I.},
  journal = {Phys. Rev. A},
  volume = {69},
  issue = {5},
  pages = {052320},
  numpages = {8},
  year = {2004},
  month = {May},
  publisher = {American Physical Society},
  doi = {10.1103/PhysRevA.69.052320},
  url = {https://link.aps.org/doi/10.1103/PhysRevA.69.052320}
}

@article{Ruppert16,
  title = {Estimation of the covariance matrix of macroscopic quantum states},
  author = {Ruppert, L\'aszl\'o and Usenko, Vladyslav C. and Filip, Radim},
  journal = {Phys. Rev. A},
  volume = {93},
  issue = {5},
  pages = {052114},
  numpages = {8},
  year = {2016},
  month = {May},
  publisher = {American Physical Society},
  doi = {10.1103/PhysRevA.93.052114},
  url = {https://link.aps.org/doi/10.1103/PhysRevA.93.052114}
}

@article{pirandola09,
  title = {Correlation matrices of two-mode bosonic systems},
  author = {Pirandola, Stefano and Serafini, Alessio and Lloyd, Seth},
  journal = {Phys. Rev. A},
  volume = {79},
  issue = {5},
  pages = {052327},
  numpages = {10},
  year = {2009},
  month = {May},
  publisher = {American Physical Society},
  doi = {10.1103/PhysRevA.79.052327},
  url = {https://link.aps.org/doi/10.1103/PhysRevA.79.052327}
}

@article{adesso04,
  title = {Determination of Continuous Variable Entanglement by Purity Measurements},
  author = {Adesso, Gerardo and Serafini, Alessio and Illuminati, Fabrizio},
  journal = {Phys. Rev. Lett.},
  volume = {92},
  issue = {8},
  pages = {087901},
  numpages = {4},
  year = {2004},
  month = {Feb},
  publisher = {American Physical Society},
  doi = {10.1103/PhysRevLett.92.087901},
  url = {https://link.aps.org/doi/10.1103/PhysRevLett.92.087901}
}

@preamble{
   "\def\Dbar{\leavevmode\lower.6ex\hbox to 0pt{\hskip-.23ex
    \accent"16\hss}D} "
}

@INPROCEEDINGS{S,
   author = {{Simon}, Sudhavathani and {Rajagopalan}, S. and {Simon}, R.},
    title = "{Indecomposable Positive Maps and a New Family of Inseparable PPT States}",
booktitle = {Quantum Computing: Back Action},
     year = {2006},
   series = {American Institute of Physics Conference Proceedings},
   volume = {864},
   editor = {{Goswami}, Debabrata},
    month = {Nov},
    pages = {67-80},
	publisher = {AIP},
      doi = {10.1063/1.2400879},
}

@book {NC,
author = "Michael A. Nielsen and Isaac L. Chuang",
title = {Quantum Computation and Quantum Information}, 
publisher = {Cambridge University Press}, 
 EDITION = {10th anniversary},
year = {2010},
isbn = {9780511976667}, 
    PAGES = {xxvi+676},
      ISBN = {0-521-63235-8; 0-521-63503-9},
   MRCLASS = {81P68 (68-01 68-02 68Q05 81-02)},
  MRNUMBER = {1796805},
url ={http://dx.doi.org/10.1017/CBO9780511976667},
       DOI = {10.1017/CBO9780511976667},
}

@preamble{
   "\def\polhk#1{\setbox0=\hbox{#1}{\ooalign{\hidewidth
    \lower1.5ex\hbox{`}\hidewidth\crcr\unhbox0}}} "
}

@article{RevModPhys.84.621,
  title = {Gaussian quantum information},
  author = {Weedbrook, Christian and Pirandola, Stefano and Garc\'ia-Patr\'on, Ra\'ul and Cerf, Nicolas J. and Ralph, Timothy C. and Shapiro, Jeffrey H. and Lloyd, Seth},
  journal = {Rev. Mod. Phys.},
  volume = {84},
  issue = {2},
  pages = {621--669},
  numpages = {0},
  year = {2012},
  month = {May},
  publisher = {American Physical Society},
  doi = {10.1103/RevModPhys.84.621},
  url = {http://link.aps.org/doi/10.1103/RevModPhys.84.621}
}

@preamble{
   "\def\cprime{$'$} "
}

@article{PhysRevLett.86.3658,
  title = {Bound Entangled Gaussian States},
  author = {Werner, R. F. and Wolf, M. M.},
  journal = {Phys. Rev. Lett.},
  volume = {86},
  issue = {16},
  pages = {3658--3661},
  year = {2001},
  month = {Apr},
  doi = {10.1103/PhysRevLett.86.3658},
  url = {http://link.aps.org/doi/10.1103/PhysRevLett.86.3658},
  publisher = {American Physical Society}
}

@article{AS,
  title = {Tensor products of convex sets and the volume of separable states on $N$ qudits},
  author = {Aubrun, Guillaume and Szarek, Stanis\l{}aw J.},
  journal = {Phys. Rev. A},
  volume = {73},
  issue = {2},
  pages = {022109},
  numpages = {10},
  year = {2006},
  month = {Feb},
  publisher = {American Physical Society},
  doi = {10.1103/PhysRevA.73.022109},
  url = {http://link.aps.org/doi/10.1103/PhysRevA.73.022109}
}

@ARTICLE{krprb3,
   author = {{Bhat}, B.~V. {Rajarama} and {Parthasarathy}, K.~R. and {Sengupta}, Ritabrata},
    title = "{On the equivalence of separability and extendability of quantum states}",
JOURNAL = {Rev. Math. Phys.},
  FJOURNAL = {Reviews in Mathematical Physics. A Journal for Both Review and
              Original Research Papers in the Field of Mathematical Physics},
    VOLUME = {29},
      YEAR = {2017},
    NUMBER = {4},
     PAGES = {1750012, 16},
      ISSN = {0129-055X},
       DOI = {10.1142/S0129055X1750012X},
       URL = {http://dx.doi.org/10.1142/S0129055X1750012X},
}

@article{haruna07,
  title = {Minimal Set of Local Measurements and Classical Communication for Two-Mode Gaussian State
Entanglement Quantification},
  author = {Haruna, Luis F. and  Oliveira, Marcos C. de and Rigolin, Gustavo},
 journal = {Phys. Rev. Lett.},
  volume = {98},
  issue = {15},
  pages = {150501},
  numpages = {4},
  year = {2007},
  month = {April},
  publisher = {American Physical Society},
  doi = {10.1103/PhysRevLett.98.150501 },
  url = {https://link.aps.org/doi/10.1103/PhysRevLett.98.150501}
  }

@article{Abiuso2021,
  title = {Measurement-Device-Independent Entanglement Detection for Continuous-Variable Systems},
  author = {Abiuso, Paolo and B\"auml, Stefan and Cavalcanti, Daniel and Ac\'{\i}n, Antonio},
  journal = {Phys. Rev. Lett.},
  volume = {126},
  issue = {19},
  pages = {190502},
  numpages = {6},
  year = {2021},
  month = {May},
  publisher = {American Physical Society},
  doi = {10.1103/PhysRevLett.126.190502},
  url = {https://link.aps.org/doi/10.1103/PhysRevLett.126.190502}
}
\section*{Appendix}
 \subsection{Measurement schemes} \label{app2}
 We first discuss schemes to obtain the information regarding the quadrature mean values as well as the covariance matrix of a two-mode Gaussian state with a limited number of measurements. From discussions in previous section, one can see  that the covariance matrix of a two-mode Gaussian state 
 can be computed by averaging   the phase space observables, e.g., $ \hat{q}_1,\,
 \hat{q}_1^2, \,  \hat{q}_2,\, \hat{q}_2^2, \,  \hat{p}_1,\, \hat{p}_1^2, \,
 \hat{p}_2,\,  \hat{p}_2^2, \, \hat{q}_1\otimes\hat{q}_2,  \,
 \hat{q}_1\otimes\hat{p}_2,  \, \hat{p}_1\otimes\hat{q}_2, \,
 \hat{p}_1\otimes\hat{p}_2, \,
 \frac{1}{2}(\hat{q}_1\hat{p}_1+\hat{p}_1\hat{q}_1)$,  and
 $\frac{1}{2}(\hat{q}_2\hat{p}_2+\hat{p}_2\hat{q}_2)$ over many copies of the
 state. For a pair of commuting operators, the corresponding observables can be
 measured jointly. 
 If we have a set
 of observables some of which are pair-wise co-measurable, then we can group them
 in such a way that the entire set can be measured with a choice from the limited
 number of measurements. We systematically describe a scheme to re-construct
 the covariance matrix of a two-mode Gaussian state with a limited number of
 measurements.
 
 Note that, $\hat{q}_1^2, \hat{q}_2^2$, and $\hat{q}_1\otimes\hat{q}_2$ are
pair-wise measurable. Also, $\hat{p}_1^2, \hat{p}_2^2$, and
$\hat{p}_1\otimes\hat{p}_2$ can be measured simultaneously. Another pair of
observables $\frac{1}{2}(\hat{q}_1\hat{p}_1+\hat{p}_1\hat{q}_1)$,  and
$\frac{1}{2}(\hat{q}_2\hat{p}_2+\hat{p}_2\hat{q}_2)$ can be co-measured. However,
$\hat{q}_1\otimes\hat{p}_2$, and $\hat{p}_1\otimes\hat{q}_2$ need to be measured
separately. Thus, the repeated measurements of  the five groups of observables
over many copies of the two-mode Gaussian state will yield complete knowledge of
the covariance matrix. 

Let's consider that $5N\, (N\gg1)$ copies of a two-mode Gaussian state
$\rho_{12}$ is shared between two parties, say Alice and Bob, where each
of them possesses one subsystem with one mode. Each of the five groups of
observables is to be measured on $N$ copies of $\rho_{12}$. The scheme goes
as follows: a) Alice measures the quadrature observable $\hat{q}_1$ on mode 1 of
$1^{\text{st}}$ $N$ copies of the shared state $\rho_{12}.$ Also, Bob
measures $\hat{q}_2$ on mode 2 of the same $N$ copies of  $\rho_{12}.$ b)
Next, Alice chooses to measure quadrature observable $\hat{p}_1$ on mode 1 of
$2^{\text{nd}}$ $N$ copies of the shared state $\rho_{12}$ and Bob measures
$\hat{p}_2$ on mode 2 of the same $N$ copies of  $\rho_{12}.$ c) Then,
Alice measures $\frac{1}{2}(\hat{q}_1\hat{p}_1+\hat{p}_1\hat{q}_1)$ on mode 1 of
$3^{\text{rd}}$ $N$ copies of the shared state $\rho_{12}.$ Bob chooses to
measure $\frac{1}{2}(\hat{q}_2\hat{p}_2+\hat{p}_2\hat{q}_2)$ on mode 2 of the
same $N$ copies of $\rho_{12}.$ d) Thereafter, Alice and Bob measure
quadrature observables $\hat{q}_1$ on mode 1 and $\hat{p}_2$ on mode 2 of
$4^{\text{th}}$ $N$ copies of $\rho_{12}$, respectively. e) At the end,
quadrature observable $\hat{p}_1$ is measured by Alice on mode 1 of the last $N$
copies of $\rho_{12},$ whereas Bob measures quadrature observable
$\hat{q}_2$ on mode 2 of the last $N$ copies of $\rho_{12}.$ Thus,
altogether five observables: $\hat{\mathcal{A}}=\hat{q}_1\otimes\hat{q}_2,
\hat{\mathcal{B}}=\hat{p}_1\otimes\hat{p}_2,
\hat{\mathcal{C}}=\frac{1}{2}(\hat{q}_1\hat{p}_1+\hat{p}_1\hat{q}_1)\otimes\frac{1}{2}(\hat{q}_2\hat{p}_2+\hat{p}_2\hat{q}_2),
\hat{\mathcal{D}}=\hat{q}_1\otimes\hat{p}_2,$ and
$\hat{\mathcal{E}}=\hat{p}_1\otimes\hat{q}_2$ are measured separately on many
copies of $\rho_{12}.$ Now, we are going to show that measurements of the
five observables are enough to compute all elements of the covariance matrix of
the two-mode Gaussian state.

Measurement outcome of $\hat{\mathcal{A}}$ will be of the form $q_1q_2,$ where
$q_1, q_2\in \mathbb{R}.$ Here, a value $q_1$ will be obtained whenever Alice
measures $\hat{q}_1$ on mode 1 of the shared state $\rho_{12}.$ Similarly,
measurement of $\hat{q}_2$ by Bob on mode 2 of $\rho_{12}$ will yield a
value $q_2.$ Also, assume that, the pair of values $(q_1, q_2)$ of quadrature
observables $(\hat{q}_1, \hat{q}_2)$ (where
$\hat{\mathcal{A}}=\hat{q}_1\otimes\hat{q}_2$)   occurs with probability $P(q_1,
q_2).$ Then, one can compute $P(q_1)$ and $P(q_2)$ as marginals of  $P(q_1,
q_2).$ Thus, by measuring $\hat{\mathcal{A}}=\hat{q}_1\otimes\hat{q}_2$ together
with finding the individual values $q_1,$ and $q_2$ one can compute respective
probabilities $P(q_1),$ and $P(q_2)$ of occurrence of these values. With the
results, one can obtain  the values $q_1^2$ and  $q_2^2$ of the respective
observables $\hat{q}^2_1,$ and  $\hat{q}^2_2)$ together with the associated
probabilities $P(q_1)$ and $P(q_2).$ Thus, measurement of
$\hat{\mathcal{A}}=\hat{q}_1\otimes\hat{q}_2$ yields triplet of values $(q_1^2,
q_2^2, q_1 q_2)$ of  observable-triplet  $(\hat{q}^2_1, \hat{q}^2_2,
\hat{\mathcal{A}}=\hat{q}_1\otimes\hat{q}_2) .$ The expectation values of the
observable-triplet are given as follows: 
\begin{eqnarray}
 \label{phys244}
 \langle \hat{q}^2_1\rangle&=&\mathrm{Tr}_{12}(\hat{q}^2_1\rho_{12})\nonumber\\
 &=& \int_{-\infty}^{+\infty} \dd q_1\int_{-\infty}^{+\infty}\dd q_2 \;q_1^2 P(\hat{q}_1=q_1, \hat{q}_2=q_2 \mid \rho_{12})\nonumber\\
 &=&\int_{-\infty}^{+\infty}\dd q_1\int_{-\infty}^{+\infty}\dd q_2 \;q_1^2 P(\hat{\mathcal{A}}=q_1q_2 \mid \rho_{12}),
 \end{eqnarray}
 \begin{eqnarray}
 \label{phys3}
 \langle \hat{q}^2_2\rangle&=&\mathrm{Tr}_{12}(\hat{q}^2_2\rho_{12})\nonumber\\
 &=&\int_{-\infty}^{+\infty}\dd q_2\int_{-\infty}^{+\infty}\dd q_1 \;q_2^2 P(\hat{\mathcal{A}}=q_1q_2 \mid \rho_{12}),
 \end{eqnarray}
 and
\begin{eqnarray}
 \label{phys4}
  \langle \hat{q}_1\otimes\hat{q}_2\rangle&=&\mathrm{Tr}_{12}(\hat{q}_1\otimes\hat{q}_2\rho_{12})\nonumber\\&=&\int_{-\infty}^{+\infty}\int_{-\infty}^{+\infty}\dd q_1 \dd q_2 \;q_1q_2 P(\hat{\mathcal{A}}=q_1q_2 \mid \rho_{12}).
 \end{eqnarray}
 Similarly,  $\langle \hat{p}^2_1\rangle,  \langle \hat{p}^2_2\rangle,$ and $ \langle \hat{p}_1\otimes\hat{p}_2\rangle$ can be obtained by measuring 
 $\hat{\mathcal{B}}=\hat{p}_1\otimes\hat{p}_2$ on $\rho_{12}.$ Again,  measurement of $\hat{\mathcal{C}}=\frac{1}{2}(\hat{q}_1\hat{p}_1+\hat{p}_1\hat{q}_1)\otimes\frac{1}{2}(\hat{q}_2\hat{p}_2+\hat{p}_2\hat{q}_2)$ leads to expectation values $\langle \frac{1}{2}(\hat{q}_1\hat{p}_1+\hat{p}_1\hat{q}_1)\rangle,$ and  $\langle \frac{1}{2}(\hat{q}_2\hat{p}_2+\hat{p}_2\hat{q}_2)\rangle.$ Finally, $\langle \hat{q}_1\otimes\hat{p}_2\rangle$ and $\langle \hat{p}_1\otimes\hat{q}_2\rangle$ can be computed by separately measuring $\hat{\mathcal{D}}=\hat{q}_1\otimes\hat{p}_2$ and $\hat{\mathcal{E}}=\hat{p}_1\otimes\hat{q}_2,$ respectively.
 
  As mentioned earlier, although the first moments of the state or expectation values of
 quadrature observables are not relevant in the detection and quantification of
 entanglement in a two-mode Gaussian state, the scheme discussed here
 can be used to obtain $\langle \hat{q}_1\rangle, \langle \hat{q}_2\rangle,
 \langle \hat{p}_1\rangle,$ and $\langle \hat{p}_2\rangle$ using  the same set of
 measurements. Precisely, measuring $\hat{\mathcal{A}}$ and $\hat{\mathcal{B}}$
 or $\hat{\mathcal{D}}$ and $\hat{\mathcal{E}}$ one can estimate $\langle
 \hat{q}_1\rangle, \langle \hat{q}_2\rangle, \langle \hat{p}_1\rangle,$ and
 $\langle \hat{p}_2\rangle.$ For example, expectation value of quadrature
 observable $\langle \hat{q}_1\rangle$ is given as follows:
 \begin{eqnarray}
 \label{phys20}
 \langle \hat{q}_1\rangle&=&\mathrm{Tr}_{12}(\hat{q}_1\rho_{12})\nonumber\\&=& \int_{-\infty}^{+\infty}\dd q_1\int_{-\infty}^{+\infty}\dd q_2 ~q_1 P(\hat{q}_1=q_1, \hat{q}_2=q_2| \rho_{12})\nonumber\\&=&\int_{-\infty}^{+\infty}\dd q_1\int_{-\infty}^{+\infty}\dd q_2 ~q_1 P(\hat{\mathcal{A}}=q_1q_2| \rho_{12}).
 \end{eqnarray}
Here, measurement outcomes of $\hat{\mathcal{A}}=\hat{q}_1\otimes\hat{q}_2,$ and
quadrature observables $\hat{q}_1, \hat{q}_1$ must be known along with the
probability of occurrence of values  $\hat{q}_1=q_1, \hat{q}_2=q_2,$ or
$\hat{\mathcal{A}}=q_1q_2,$ i.e., $P(\hat{q}_1=q_1, \hat{q}_2=q_2|
\rho_{12})$ or $P(\hat{\mathcal{A}}=q_1q_2| \rho_{12}).$ Applying the
similar method one obtains the first moments of the state. Thus, interestingly, the set of {\em five} measurements is
sufficient to reconstruct the covariance matrix of a  two-mode Gaussian state. Note that the aforesaid method provides the mean values $\langle\hat{q}_1\rangle, \langle\hat{p}_1\rangle, \langle\hat{q}_2\rangle,$ and $\langle\hat{p}_2\rangle$ as well as the covariance matrix for any two-mode state, and not necessarily only for two-mode Gaussian states. As a result, this method gives rise to tomography of any two-mode Gaussian states using measurements on the individual modes.

Next, we present another scheme to estimate the elements of the covariance matrix with the same set of five measurements. It will be interesting to find a set of measurable quantities to experimentally realize the set of five measurements, e.g., using homodyne measurements. 

 In the second scheme, Alice and Bob divide the $5N(N \gg1)$ copies of two-mode Gaussian state $\rho_{12}$ in two groups, e.g., a group with $4N$ copies of $\rho_{12}$ and another one with $N$ copies of $\rho_{12}.$ Here, the exchange of 1 bit of classical communication  between Alice and Bob is required to distinguish between two groups. Let's consider, at first Alice and Bob choose to perform measurements on $4N$ copies of $\rho_{12}.$ Thereafter, Alice randomly measures $\hat{q}_1$ and $\hat{p}_1$ on mode 1, whereas Bob randomly measures $\hat{q}_2$ and $\hat{p}_2$ on mode 2 of $\rho_{12}$ from the group with $4N$ copies of the state. Note that, effectively, Alice and Bob measure $\hat{\mathcal{A}}=\hat{q}_1\otimes\hat{q}_2,\, \hat{\mathcal{B}}=\hat{p}_1\otimes\hat{p}_2, \,\hat{\mathcal{D}}=\hat{q}_1\otimes\hat{p}_2,$ and $\hat{\mathcal{E}}=\hat{p}_1\otimes\hat{q}_2$ at random. Classical communication between Alice and Bob is necessary to know precisely if $\hat{q}_1$ is measured along with $\hat{q}_2$ or $\hat{p}_2$ and $\hat{p}_1$ is measured along with $\hat{q}_2$ or $\hat{p}_2.$ In turn, this will lead to joint probabilities $P(q_1, q_2), \, P(q_1,p _2), \, P(p_1, q_2),$ and $P(p_1, p_2).$ On each of the remaining $N$ copies of $\rho_{12}$ Alice and Bob measure 
$\frac{1}{2}(\hat{q}_1\hat{p}_1+\hat{p}_1\hat{q}_1),$ and $\frac{1}{2}(\hat{q}_2\hat{p}_2+\hat{p}_2\hat{q}_2),$ respectively. This is equivalent to measure~$\hat{\mathcal{C}}=\frac{1}{2}(\hat{q}_1\hat{p}_1+\hat{p}_1\hat{q}_1)\otimes\frac{1}{2}(\hat{q}_2\hat{p}_2+\hat{p}_2\hat{q}_2)$. In this case, no classical communication between Alice and Bob is needed to compute joint probability $P\left(\frac{1}{2}(\hat{q}_1\hat{p}_1+\hat{p}_1\hat{q}_1), \frac{1}{2}(\hat{q}_2\hat{p}_2+\hat{p}_2\hat{q}_2)\right).$ 

Interestingly, Alice can estimate expectation values  $\langle
\hat{q}^2_1\rangle,$ and $\langle \hat{p}^2_1\rangle$ by separately measuring
$\hat{q}_1,$ and  $\hat{p}_1$ on many copies of mode 1 of $\rho_{12}.$
Likewise, Bob obtains expectation values  $\langle \hat{q}^2_2\rangle,$ and
$\langle \hat{p}^2_2\rangle$ by separately performing measurements of
$\hat{q}_2,$ and  $\hat{p}_2$ on many copies of mode 2 of $\rho_{12}.$ Note
that, expectation values which are defining correlations between measurements on
two modes e.g., $\langle \hat{q}_1\otimes\hat{q}_2\rangle, \,\langle
\hat{p}_1\otimes\hat{p}_2\rangle, \, \langle
\frac{1}{2}(\hat{q}_1\hat{p}_1+\hat{p}_1\hat{q}_1)\otimes\frac{1}{2}(\hat{q}_2\hat{p}_2+\hat{p}_2\hat{q}_2)\rangle,
\, \langle \hat{q}_1\otimes\hat{p}_2\rangle,$ and $\langle
\hat{p}_1\otimes\hat{q}_2\rangle$ are computed by separately measuring
$\hat{\mathcal{A}}=\hat{q}_1\otimes\hat{q}_2, \,
\hat{\mathcal{B}}=\hat{p}_1\otimes\hat{p}_2, \,
\hat{\mathcal{C}}=\frac{1}{2}(\hat{q}_1\hat{p}_1+\hat{p}_1\hat{q}_1)\otimes\frac{1}{2}(\hat{q}_2\hat{p}_2+\hat{p}_2\hat{q}_2),
\,\hat{\mathcal{D}}=\hat{q}_1\otimes\hat{p}_2,$ and
$\hat{\mathcal{E}}=\hat{p}_1\otimes\hat{q}_2$ on many copies of
$\rho_{12}.$ The first moments of the state are estimated in Section~\ref{s2}. Thus, all the elements of the covariance matrix of a two-mode
Gaussian state are estimated with less resources than the full state tomography. 
   

\subsection{Calculations of elements of the covariance matrix of a displaced squeezed thermal state} 
\label{reference}
In our analysis, we have considered the displaced squeezed thermal state as a reference state. The thermal state is represented by a density matrix of the following form:
 \begin{equation}
 \label{append1}
 \rho_r=\sum_{n=0}^{\infty}\frac{\overline{n}_r^n}{(\overline{n}_r+1)^{n+1}}|n\rangle\langle n|,
 \end{equation}
where $\rho_r$ is expressed in the number state basis and $\bar{n}_r$ is
the mean photon number of the thermal state. Displaced squeezed thermal state
$\rho^{DS}_r$ is obtained after applying a unitary transformation
$\hat{U}_r=\hat{D}(\alpha_r)\hat{S}(\xi_r)$ on the state $\rho_r,$ where
$\hat{D}(\alpha_r)$ is a displacement operator:
$\hat{D}(\alpha_r)=\exp\left(\alpha_r\hat{a}_r^{\dagger}-\alpha_r^*\hat{a}_r\right),$
and $\hat{S}(\xi_r)$ is a squeezing operator: $\hat{S}(\xi_r)=\exp\frac{1}{2}
\left(\xi_r^*\hat{a}_r^2-\xi_r\hat{a}^{\dagger2}_r\right).$ Here
$\alpha_r=d_re^{\imath\beta_r},$ and $\xi_r=\theta_re^{\imath\gamma_r}$. Note
that, for an unbiased displaced squeezed thermal state $\beta_r=\gamma_r=0.$ A
biased $\rho^{DS}_r$ is defined as follows:
\begin{equation}
 \label{append2}
\rho^{DS}_r=\hat{D}(\alpha_r)\hat{S}(\xi_r)\rho_r\hat{S}(\xi_r)^{\dagger}\hat{D}(\alpha_r)^{\dagger}.
 \end{equation}
Thereafter, the expectation value of an observable $\hat{\mathcal{O}}$ reads
\begin{eqnarray}
 \label{append3}
 \langle \hat{\mathcal{O}} \rangle &=& \tr \left(\rho^{DS}_r\hat{\mathcal{O}} \right)=\tr \left(\hat{D}(\alpha_r)\hat{S}(\xi_r)\rho_r\hat{S}(\xi_r)^{\dagger}\hat{D}(\alpha_r)^{\dagger}\hat{\mathcal{O}} \right)\nonumber\\
&=& \tr \left(\rho_r\hat{S}(\xi_r)^{\dagger}\hat{D}(\alpha_r)^{\dagger}\hat{\mathcal{O}}\hat{D}(\alpha_r)\hat{S}(\xi_r)\right),
 \end{eqnarray}
 where the last equality follows from the invariance of trace under cyclic permutation of operators. Next, in our derivation of expectation values of observables, we consider the following set of transformations of annihilation and creation operators:
\begin{eqnarray}
 \label{append4}
\hat{a}_r &\to& \hat{a}_r\cosh \theta_r-e^{\imath\gamma_r}\hat{a}_r^{\dagger}\sinh \theta_r+\alpha_r\nonumber\\
\hat{a}_r^{\dagger} &\to& \hat{a}_r^{\dagger}\cosh \theta_r-e^{-\imath\gamma_r}\hat{a}_r\sinh \theta_r+\alpha^*_r.
 \end{eqnarray}
 
 For a displaced squeezed thermal state, the expectation values of quadrature observables and the square of the quadrature observables read
\begin{eqnarray}
 \label{append5}
\langle \hat{q}_r\rangle &=& \sqrt{2}d_r\cos \beta_r, \quad \langle \hat{p}_r\rangle=\sqrt{2}d_r\sin \beta_r\nonumber\\
\langle \hat{q}^2_r\rangle &=& \left( \overline{n}_r+\frac{1}{2} \right) \left(e^{2\theta_r}\sin^2\frac{\gamma_r}{2}+e^{-2\theta_r}\cos^2\frac{\gamma_r}{2} \right)+2d_r^2\cos^2\beta_r,\nonumber\\
\langle \hat{p}^2_r\rangle &=& \left(\overline{n}_r+\frac{1}{2}\right) \left(e^{2\theta_r}\cos^2\frac{\gamma_r}{2}+e^{-2\theta_r}\sin^2\frac{\gamma_r}{2} \right)+2d_r^2\sin^2\beta_r.
 \end{eqnarray}
 In addition, we compute 
\begin{eqnarray}
 \label{append6}
\langle \hat{q}_r\hat{p}_r\rangle &=&d_r^2\sin 2\beta_r- \left(\overline{n}_r+\frac{1}{2}\right) \sinh 2\theta_r\sin\gamma_r+\frac{\imath}{2}\nonumber\\
\langle \hat{p}_r\hat{q}_r\rangle &=& d_r^2\sin 2\beta_r- \left(\overline{n}_r+\frac{1}{2} \right)\sinh 2 \theta_r\sin\gamma_r-\frac{\imath}{2}.
 \end{eqnarray}
 Note that, for unbiased displaced squeezed thermal state  $\left\langle \hat{q}_r\hat{p}_r \right\rangle=- \left\langle \hat{p}_r\hat{q}_r \right\rangle= \frac{\imath}{2}.$ Similarly, one can choose  $d_r, \overline{n}_r, \beta_r, \gamma_r,$ and $\theta_r$ in such a way that, we obtain $\left\langle \hat{q}_r\hat{p}_r \right\rangle=-\left\langle \hat{p}_r\hat{q}_r \right\rangle=\frac{\imath}{2},$ for $d_r^2\sin 2\beta_r= \left(\overline{n}_r+\frac{1}{2}\right)\sinh 2\theta_r\sin\gamma_r,$ where $\beta_r,$ $\gamma_r,$ and $\theta_r$ are non-zero.
From $\eqref{append5},$ one can calculate
\begin{eqnarray}
 \label{append7}
\langle \hat{q}^2_r\rangle-\langle\hat{p}^2_r\rangle &=& 2d_r^2\cos 2\beta_r-(2\overline{n}_r+1)\sinh 2\theta_r\cos\gamma_r\nonumber\\
\langle \hat{q}^2_r\rangle+\langle\hat{p}^2_r\rangle &=& 2d_r^2+(2\overline{n}_r+1)\cosh 2\theta_r.
 \end{eqnarray}
The mean photon number in a single mode of a displaced squeezed thermal state is given by
\begin{eqnarray}
\label{total}
n_r^{(t)}=d_r^2+n_r\cosh 2\theta_r+\sinh^2\theta_r.
\end{eqnarray}

\subsection{Robustness Analysis of the Proposed Scheme}
\label{robustness}

 Considering the photon number difference of the incident two-mode Gaussian state at the two outputs of the beam splitter (see Fig.~\ref{figg10}(a)), one can estimate the expectation value of $\hat{S}_1'(\phi),$
 i.e.,
\begin{eqnarray}
\label{stokes0000}
\langle\hat{S'}_1(\phi)\rangle = \langle\tilde{I}'_{2}-\tilde{I}'_{1}\rangle = \langle\hat{\tilde{a}}_2'^{\dagger}\hat{\tilde{a}}_2'-\hat{\tilde{a}}_1'^{\dagger}\hat{\tilde{a}}_1'\rangle = \eta\langle\hat{S}_1(\phi)\rangle.
\end{eqnarray}
Following $\eqref{stokes00},$ for a fixed reference state, we can write
\begin{eqnarray}
\label{stokes00n}
\langle\hat{S'}_1(\phi)\rangle=\langle\hat{q'}_k\rangle\langle\hat{q}_{r_k}^{\phi}\rangle+\langle\hat{p'}_k\rangle\langle\hat{p}_{r_k}^{\phi}\rangle.
 \end{eqnarray}
 Here, $\langle\hat{q'}_k\rangle,$ and $\langle\hat{p'}_k\rangle$ are the first moments of the signal state after the photon loss, whereas $\langle\hat{q}_{r_k}^{\phi}\rangle,$ and $\langle\hat{p}_{r_k}^{\phi}\rangle$ are associated with the reference state. We can estimate $\langle\hat{q'}_k\rangle,$ and $\langle\hat{p'}_k\rangle$ in terms of $\langle\hat{q}_k\rangle,$ and $\langle\hat{p}_k\rangle$ by solving equation $\eqref{stokes00n},$
for $\phi=0,$ and $\phi=\frac{\pi}{2}.$ 

 
 Similarly, we can estimate $\langle\hat{q'}_k^2\rangle, \langle\hat{p'}_k^2\rangle,$ $\langle\frac{1}{2}(\hat{q'}_k\hat{p'}_k+\hat{p'}_k\hat{q'}_k)\rangle$ from
\begin{eqnarray}
\label{stokes0010}
&\langle\hat{S'}_1^2(\phi)\rangle=\langle(\tilde{I}'_{2}-\tilde{I}'_{1})^2\rangle
&\nonumber\\
&=\eta^2\langle\hat{S}_1^2(\phi)\rangle+\eta(1-\eta)\langle\hat{S}_0^{(k)}\rangle,& 
 \end{eqnarray}
 where $\langle\hat{S}_0^{(k)}\rangle$ is the total  photon number at the two detected output ports in Fig.~\ref{figg10} (b), and \ref{figg10}(c) for the k-th signal mode, i.e., $\hat{S}_0^{(k)}=\hat{\tilde{a}}_2^{\dagger}\hat{\tilde{a}}_2+\hat{\tilde{a}}_1^{\dagger}\hat{\tilde{a}}_1,$ where \(\hat{\tilde{a}}_1\), and \(\hat{\tilde{a}}_2\) are defined in Eq.~\eqref{single}.
 Precisely, we obtain 
   \begin{eqnarray}
\label{stokes0011}
&\langle\hat{q'}_k^2\rangle=\eta^2\langle\hat{q}_k^2\rangle+\frac{\eta(1-\eta)}{\langle\hat{q}_{r_k}^2\rangle+\langle\hat{p}_{r_k}^2\rangle}\langle\hat{S}_0^{(k)}\rangle+\frac{1-\eta^2}{2(\langle\hat{q}_{r_k}^2\rangle+\langle\hat{p}_{r_k}^2\rangle)}
&\nonumber\\
&\langle\hat{p'}_k^2\rangle=\eta^2\langle\hat{p}_k^2\rangle+\frac{\eta(1-\eta)}{\langle\hat{q}_{r_k}^2\rangle+\langle\hat{p}_{r_k}^2\rangle}\langle\hat{S}_0^{(k)}\rangle+\frac{1-\eta^2}{2(\langle\hat{q}_{r_k}^2\rangle+\langle\hat{p}_{r_k}^2\rangle)},
&\nonumber
\end{eqnarray}
and
\begin{eqnarray}
&\langle\frac{1}{2}(\hat{q'}_k\hat{p'}_k+\hat{p'}_k\hat{q}_k)\rangle=\eta^2\langle\frac{1}{2}(\hat{q}_k\hat{p}_k+\hat{p}_k\hat{q}_k)\rangle.&
 \end{eqnarray}

 The diagonal elements of the matrix $C$ are obtained from the second moments $\langle\hat{S}_1'^{2}(\phi_1=0)\rangle,$ and $\langle\hat{S}_1'^{2}(\phi_2=0)\rangle.$ Specifically, we have 
\begin{eqnarray}
\label{stokes002n}
&\langle\hat{S}_1'^{2}(\phi_1=0)\rangle=\langle(I'_{4}-I'_{3})^2\rangle=\langle(\hat{a}_4'^{\dagger}\hat{a}_4'-\hat{a}_3'^{\dagger}\hat{a}_3')^2\rangle&\nonumber\\&
=\eta^2\langle\hat{S}_1^{2}(\phi_1=0)\rangle+\eta(1-\eta)\langle\hat{S}_0^{(34)}\rangle,&
 \end{eqnarray}
 where $\langle\hat{S}_0^{(34)}\rangle=\langle\hat{a}_4^{\dagger}\hat{a}_4+\hat{a}_3^{\dagger}\hat{a}_3\rangle$ denotes total number of photons in mode 3 and mode 4. Similarly, 
\begin{eqnarray}
\label{stokes004n}
&\langle\hat{S}_1'^{2}(\phi_2=0)\rangle=\langle(I'_{6}-I'_{5})^2\rangle=\langle(\hat{a}_6'^{\dagger}\hat{a}_6'-\hat{a}_5'^{\dagger}\hat{a}_5')^2\rangle&\nonumber\\
&=\eta^2\langle\hat{S}_1^{2}(\phi_2=0)\rangle+\eta(1-\eta)\langle\hat{S}_0^{(56)}\rangle,&
 \end{eqnarray}
 where $\langle\hat{S}_0^{(56)}\rangle=\langle\hat{a}_6^{\dagger}\hat{a}_6+\hat{a}_5^{\dagger}\hat{a}_5\rangle.$ Similarly,  
we estimate 
\begin{eqnarray}
\label{stokes00121}
&\langle\hat{q'}_1\otimes \hat{q'}_2\rangle=\eta^2\langle\hat{q}_1\otimes \hat{q}_2\rangle+\frac{\eta(1-\eta)}{\langle\hat{q}_{d}^2\rangle\langle\hat{p}_{c}^2\rangle-\langle\hat{q}_{c}^2\rangle\langle\hat{p}_{d}^2\rangle}(\langle\hat{p}_{d}^2\rangle\langle\hat{S}_{0}^{(34)}\rangle+\langle\hat{p}_{c}^2\rangle\langle\hat{S}_{0}^{(56)}\rangle)+\frac{1-\eta^2}{2(\langle\hat{q}_{d}^2\rangle\langle\hat{p}_{c}^2\rangle-\langle\hat{q}_{c}^2\rangle\langle\hat{p}_{d}^2\rangle)}(\langle\hat{p}_{c}^2\rangle+\langle\hat{p}_{d}^2\rangle)
&\nonumber\\
&-\frac{1}{2}(\frac{\langle\hat{q}_{d}^2\rangle\langle\hat{p}_{c}^2\rangle+\langle\hat{q}_{c}^2\rangle\langle\hat{p}_{d}^2\rangle+2\langle\hat{p}_{c}^2\rangle\langle\hat{p}_{d}^2\rangle}{\langle\hat{q}_{d}^2\rangle\langle\hat{p}_{c}^2\rangle-\langle\hat{q}_{c}^2\rangle\langle\hat{p}_{d}^2\rangle})(\frac{\eta(1-\eta)}{\langle\hat{q}_{r_1}^2\rangle+\langle\hat{p}_{r_1}^2\rangle}\langle\hat{S}_{0}^{(1)}\rangle+\frac{1-\eta^2}{2(\langle\hat{q}_{r_1}^2\rangle+\langle\hat{p}_{r_1}^2\rangle)}+\frac{\eta(1-\eta)}{\langle\hat{q}_{r_2}^2\rangle+\langle\hat{p}_{r_2}^2\rangle}\langle\hat{S}_{0}^{(2)}\rangle+\frac{1-\eta^2}{2(\langle\hat{q}_{r_2}^2\rangle+\langle\hat{p}_{r_2}^2\rangle)}),&
\end{eqnarray}
and
 \begin{eqnarray}
\label{stokes00122}
&\langle\hat{p'}_1\otimes \hat{p'}_2\rangle=\eta^2\langle\hat{p}_1\otimes \hat{p}_2\rangle-\frac{\eta(1-\eta)}{\langle\hat{q}_{d}^2\rangle\langle\hat{p}_{c}^2\rangle-\langle\hat{q}_{c}^2\rangle\langle\hat{p}_{d}^2\rangle}(\langle\hat{q}_{d}^2\rangle\langle\hat{S}_{0}^{(34)}\rangle+\langle\hat{q}_{c}^2\rangle\langle\hat{S}_{0}^{(56)}\rangle)-\frac{1-\eta^2}{2(\langle\hat{q}_{d}^2\rangle\langle\hat{p}_{c}^2\rangle-\langle\hat{q}_{c}^2\rangle\langle\hat{p}_{d}^2\rangle)}(\langle\hat{q}_{c}^2\rangle+\langle\hat{q}_{d}^2\rangle)
&\nonumber\\
&+\frac{1}{2}(\frac{\langle\hat{q}_{d}^2\rangle\langle\hat{p}_{c}^2\rangle+\langle\hat{q}_{c}^2\rangle\langle\hat{p}_{d}^2\rangle+2\langle\hat{q}_{c}^2\rangle\langle\hat{q}_{d}^2\rangle}{\langle\hat{q}_{d}^2\rangle\langle\hat{p}_{c}^2\rangle-\langle\hat{q}_{c}^2\rangle\langle\hat{p}_{d}^2\rangle})(\frac{\eta(1-\eta)}{\langle\hat{q}_{r_1}^2\rangle+\langle\hat{p}_{r_1}^2\rangle}\langle\hat{S}_{0}^{(1)}\rangle+\frac{1-\eta^2}{2(\langle\hat{q}_{r_1}^2\rangle+\langle\hat{p}_{r_1}^2\rangle)}+\frac{\eta(1-\eta)}{\langle\hat{q}_{r_2}^2\rangle+\langle\hat{p}_{r_2}^2\rangle}\langle\hat{S}_{0}^{(2)}\rangle+\frac{1-\eta^2}{2(\langle\hat{q}_{r_2}^2\rangle+\langle\hat{p}_{r_2}^2\rangle)}).&
\end{eqnarray}

Following the previous arguments, the off-diagonal elements of the matrix $C$ are obtained from the measurements $\langle\hat{S}_1'^{2}(\phi_2=\frac{\pi}{4})\rangle$ and $\langle\hat{S}_3'(\phi_1=0, \phi_2=0)\rangle=\eta\langle\hat{S}_3(\phi_1=0, \phi_2=0)\rangle.$ For simplicity, we assume the first order moments vanish, i.e., $\langle\hat{q}_1\rangle=\langle\hat{q}_2\rangle=\langle\hat{p}_1\rangle=\langle\hat{p}_2\rangle=0$~\footnote{Although this assumption looses our primary goal of `universal' detection of entanglement in two-mode Gaussian states ({\it i.e.}, whether or not they have zero local means), nevertheless, this assumption has been made here just to have relatively simpler expressions, and the same procedure can, in fact, be carried forward without this assumption.}. Under these assumptions,  we compute
  \begin{eqnarray}
\label{stokes00123}
&\langle\hat{q'}_1\otimes \hat{p'}_2\rangle=\frac{\eta(1+\eta)}{2}\langle\hat{q}_1\otimes \hat{p}_2\rangle-\frac{\eta(1-\eta)}{2}\langle\hat{p}_1\otimes \hat{q}_2\rangle-\frac{(1-\eta)}{2}(\langle\hat{q}_d\rangle\langle \hat{p}_c\rangle-\langle\hat{p}_d\rangle\langle \hat{q}_c\rangle)+\frac{\frac{1}{2}(1-\eta^2)+\eta(1-\eta)\langle\hat{S}_{0}^{(56)}\rangle}{\langle\hat{q}_{d}^2\rangle-\langle\hat{p}_{d}^2\rangle}
&\nonumber\\
&+\eta(1-\eta)\frac{(\langle\hat{q}_{d}^2\rangle+\langle\hat{p}_{d}^2\rangle)\langle\hat{S}_{0}^{(34)}\rangle+\frac{\langle\hat{q}_{d}^2\rangle+\langle\hat{p}_{d}^2\rangle}{\langle\hat{q}_{d}^2\rangle-\langle\hat{p}_{d}^2\rangle}(\langle\hat{q}_{c}^2\rangle-\langle\hat{p}_{c}^2\rangle)\langle\hat{S}_{0}^{(56)}\rangle}{2(\langle\hat{q}_{d}^2\rangle\langle\hat{p}_{c}^2\rangle-\langle\hat{q}_{c}^2\rangle\langle\hat{p}_{d}^2\rangle)}+\frac{1-\eta^2}{4}(\frac{\langle\hat{q}_{d}^2\rangle+\langle\hat{p}_{d}^2\rangle}{\langle\hat{q}_{d}^2\rangle\langle\hat{p}_{c}^2\rangle-\langle\hat{q}_{c}^2\rangle\langle\hat{p}_{d}^2\rangle}+\frac{\langle\hat{q}_{c}^2\rangle\langle\hat{q}_{d}^2\rangle-\langle\hat{p}_{c}^2\rangle\langle\hat{p}_{d}^2\rangle}{\langle\hat{q}_{d}^2\rangle\langle\hat{p}_{c}^2\rangle-\langle\hat{q}_{c}^2\rangle\langle\hat{p}_{d}^2\rangle}-1)
&\nonumber\\
&-\frac{(\langle\hat{q}_{c}^2\rangle+\langle\hat{p}_{c}^2\rangle)(\langle\hat{q}_{d}^2\rangle+\langle\hat{p}_{d}^2\rangle)}{2(\langle\hat{q}_{d}^2\rangle\langle\hat{p}_{c}^2\rangle-\langle\hat{q}_{c}^2\rangle\langle\hat{p}_{d}^2\rangle)}(\frac{\eta(1-\eta)}{\langle\hat{q}_{r_1}^2\rangle+\langle\hat{p}_{r_1}^2\rangle}\langle\hat{S}_{0}^{(1)}\rangle+\frac{1-\eta^2}{2(\langle\hat{q}_{r_1}^2\rangle+\langle\hat{p}_{r_1}^2\rangle)}+\frac{\eta(1-\eta)}{\langle\hat{q}_{r_2}^2\rangle+\langle\hat{p}_{r_2}^2\rangle}\langle\hat{S}_{0}^{(2)}\rangle+\frac{1-\eta^2}{2(\langle\hat{q}_{r_2}^2\rangle+\langle\hat{p}_{r_2}^2\rangle)}),&
\end{eqnarray}
and
\begin{eqnarray}
&\langle\hat{p'}_1\otimes \hat{q'}_2\rangle=\frac{\eta(1+\eta)}{2}\langle\hat{p}_1\otimes \hat{q}_2\rangle-\frac{\eta(1-\eta)}{2}\langle\hat{q}_1\otimes \hat{p}_2\rangle+\frac{(1-\eta)}{2}(\langle\hat{q}_d\rangle\langle \hat{p}_c\rangle-\langle\hat{p}_d\rangle\langle \hat{q}_c\rangle)+\frac{\frac{1}{2}(1-\eta^2)+\eta(1-\eta)\langle\hat{S}_{0}^{(56)}\rangle}{\langle\hat{q}_{d}^2\rangle-\langle\hat{p}_{d}^2\rangle}
&\nonumber\\
&+\eta(1-\eta)\frac{(\langle\hat{q}_{d}^2\rangle+\langle\hat{p}_{d}^2\rangle)\langle\hat{S}_{0}^{(34)}\rangle+\frac{\langle\hat{q}_{d}^2\rangle+\langle\hat{p}_{d}^2\rangle}{\langle\hat{q}_{d}^2\rangle-\langle\hat{p}_{d}^2\rangle}(\langle\hat{q}_{c}^2\rangle-\langle\hat{p}_{c}^2\rangle)\langle\hat{S}_{0}^{(56)}\rangle}{2(\langle\hat{q}_{d}^2\rangle\langle\hat{p}_{c}^2\rangle-\langle\hat{q}_{c}^2\rangle\langle\hat{p}_{d}^2\rangle)} +\frac{1-\eta^2}{4}(\frac{\langle\hat{q}_{d}^2\rangle+\langle\hat{p}_{d}^2\rangle}{\langle\hat{q}_{d}^2\rangle\langle\hat{p}_{c}^2\rangle-\langle\hat{q}_{c}^2\rangle\langle\hat{p}_{d}^2\rangle}+\frac{\langle\hat{q}_{c}^2\rangle\langle\hat{q}_{d}^2\rangle-\langle\hat{p}_{c}^2\rangle\langle\hat{p}_{d}^2\rangle}{\langle\hat{q}_{d}^2\rangle\langle\hat{p}_{c}^2\rangle-\langle\hat{q}_{c}^2\rangle\langle\hat{p}_{d}^2\rangle}-1)
&\nonumber\\
&-\frac{(\langle\hat{q}_{c}^2\rangle+\langle\hat{p}_{c}^2\rangle)(\langle\hat{q}_{d}^2\rangle+\langle\hat{p}_{d}^2\rangle)}{2(\langle\hat{q}_{d}^2\rangle\langle\hat{p}_{c}^2\rangle-\langle\hat{q}_{c}^2\rangle\langle\hat{p}_{d}^2\rangle)}(\frac{\eta(1-\eta)}{\langle\hat{q}_{r_1}^2\rangle+\langle\hat{p}_{r_1}^2\rangle}\langle\hat{S}_{0}^{(1)}\rangle+\frac{1-\eta^2}{2(\langle\hat{q}_{r_1}^2\rangle+\langle\hat{p}_{r_1}^2\rangle)}+\frac{\eta(1-\eta)}{\langle\hat{q}_{r_2}^2\rangle+\langle\hat{p}_{r_2}^2\rangle}\langle\hat{S}_{0}^{(2)}\rangle+\frac{1-\eta^2}{2(\langle\hat{q}_{r_2}^2\rangle+\langle\hat{p}_{r_2}^2\rangle)}).&
\end{eqnarray}
We can write the elements of the covariance matrix in a concise manner. From Eqs.~(\ref{stokes0011}), (\ref{stokes00121}), (\ref{stokes00122}), (\ref{stokes00123}), the primed single-mode second moments are
\begin{align}
\langle \hat q_k'^2\rangle &= 
\eta^2 \langle \hat q_k^2\rangle
+ \eta(1-\eta)\,\Sigma_{r_k}\,\langle \hat S_0^{(k)}\rangle
+ \frac{1-\eta^2}{2}\Sigma_{r_k},\\
\langle \hat p_k'^2\rangle &= 
\eta^2 \langle \hat p_k^2\rangle
+ \eta(1-\eta)\,\Sigma_{r_k}\,\langle \hat S_0^{(k)}\rangle
+ \frac{1-\eta^2}{2}\Sigma_{r_k},\\
&\frac12\langle \hat q_k'\hat p_k' + \hat p_k'\hat q_k'\rangle= \eta^2\,\frac12\langle \hat q_k\hat p_k + \hat p_k\hat q_k\rangle,
\end{align}
where
\[
\Sigma_{r_k} := 
\frac{1}{\langle \hat q_{r_k}^2\rangle + \langle \hat p_{r_k}^2\rangle}.
\]

If $C'$ represents the cross-correlation block of the covariance matrix for a two-mode system after accounting for photon loss, then the diagonal entries of $C'$ are
\begin{align}
\langle \hat q_1'\otimes\hat q_2'\rangle
&=
\eta^2\langle \hat q_1\otimes\hat q_2\rangle
+ \frac{\eta(1-\eta)}{D_{cd}}\left(
\langle \hat p_d^2\rangle\langle \hat S_0^{(34)}\rangle
+ \langle \hat p_c^2\rangle\langle \hat S_0^{(56)}\rangle\right)
\nonumber\\
&\quad
+ \frac{1-\eta^2}{2 D_{cd}}(\langle \hat p_c^2\rangle+\langle \hat p_d^2\rangle)
-\frac12\frac{
\langle \hat q_d^2\rangle\langle \hat p_c^2\rangle
+ \langle \hat q_c^2\rangle\langle \hat p_d^2\rangle
+2\langle \hat p_c^2\rangle\langle \hat p_d^2\rangle}{D_{cd}}
\,F_{\rm ref},
\\[1em]
\langle \hat p_1'\otimes\hat p_2'\rangle
&=
\eta^2\langle \hat p_1\otimes\hat p_2\rangle
-\frac{\eta(1-\eta)}{D_{cd}}\left(
\langle \hat q_d^2\rangle\langle \hat S_0^{(34)}\rangle
+ \langle \hat q_c^2\rangle\langle \hat S_0^{(56)}\rangle\right)
\nonumber\\
&\quad
-\frac{1-\eta^2}{2 D_{cd}}(\langle \hat q_c^2\rangle+\langle \hat q_d^2\rangle)
+\frac12\frac{
\langle \hat q_d^2\rangle\langle \hat p_c^2\rangle
+ \langle \hat q_c^2\rangle\langle \hat p_d^2\rangle
+2\langle \hat q_c^2\rangle\langle \hat q_d^2\rangle}{D_{cd}}
\,F_{\rm ref}.
\end{align}

For the off-diagonal entries (with vanishing first moments of the signal):
\begin{align}
\langle \hat q_1'\otimes\hat p_2'\rangle
&=
-\frac{1-\eta}{2}
(\langle \hat q_d\rangle\langle \hat p_c\rangle
-\langle \hat p_d\rangle\langle \hat q_c\rangle)
+\frac{\frac{1}{2}(1-\eta^2)
+\eta(1-\eta)\langle \hat S_0^{(56)}\rangle}{\langle \hat q_d^2\rangle-\langle \hat p_d^2\rangle}
\nonumber\\
&\quad
+\eta(1-\eta)
\frac{(\langle \hat q_d^2\rangle+\langle \hat p_d^2\rangle)\langle \hat S_0^{(34)}\rangle
+(\langle \hat q_d^2\rangle+\langle \hat p_d^2\rangle)
\left(\frac{\langle \hat q_c^2\rangle-\langle \hat p_c^2\rangle}{\langle \hat q_d^2\rangle-\langle \hat p_d^2\rangle}\right)
\langle \hat S_0^{(56)}\rangle}{2D_{cd}}
\nonumber\\
&\quad
+\frac{1-\eta^2}{4}\left(
\frac{\langle \hat q_d^2\rangle+\langle \hat p_d^2\rangle}{D_{cd}}
+\frac{\langle \hat q_c^2\rangle\langle \hat q_d^2\rangle
-\langle \hat p_c^2\rangle\langle \hat p_d^2\rangle}{D_{cd}}
-1\right)
\nonumber\\
&\quad
-\frac{(\langle \hat q_c^2\rangle+\langle \hat p_c^2\rangle)
(\langle \hat q_d^2\rangle+\langle \hat p_d^2\rangle)}{2D_{cd}}
\,F_{\rm ref}.
\end{align}

\begin{align}
\langle \hat p_1'\otimes\hat q_2'\rangle
&=
\frac{1-\eta}{2}
(\langle \hat q_d\rangle\langle \hat p_c\rangle
-\langle \hat p_d\rangle\langle \hat q_c\rangle)
+\frac{\frac{1}{2}(1-\eta^2)
+\eta(1-\eta)\langle \hat S_0^{(56)}\rangle}{\langle \hat q_d^2\rangle-\langle \hat p_d^2\rangle}
\nonumber\\
&\quad
+\eta(1-\eta)
\frac{(\langle \hat q_d^2\rangle+\langle \hat p_d^2\rangle)\langle \hat S_0^{(34)}\rangle
+(\langle \hat q_d^2\rangle+\langle \hat p_d^2\rangle)
\left(\frac{\langle \hat q_c^2\rangle-\langle \hat p_c^2\rangle}{\langle \hat q_d^2\rangle-\langle \hat p_d^2\rangle}\right)
\langle \hat S_0^{(56)}\rangle}{2D_{cd}}
\nonumber\\
&\quad
+\frac{1-\eta^2}{4}\left(
\frac{\langle \hat q_d^2\rangle+\langle \hat p_d^2\rangle}{D_{cd}}
+\frac{\langle \hat q_c^2\rangle\langle \hat q_d^2\rangle
-\langle \hat p_c^2\rangle\langle \hat p_d^2\rangle}{D_{cd}}
-1\right)
\nonumber\\
&\quad
-\frac{(\langle \hat q_c^2\rangle+\langle \hat p_c^2\rangle)
(\langle \hat q_d^2\rangle+\langle \hat p_d^2\rangle)}{2D_{cd}}
\,F_{\rm ref},
\end{align}
where
\begin{equation}
  D_{cd}
  := \langle\hat q_d^2\rangle\langle\hat p_c^2\rangle
   - \langle\hat q_c^2\rangle\langle\hat p_d^2\rangle,
\end{equation}
and
\begin{equation}
F_{\rm ref} :=
\eta(1-\eta)\Sigma_{r_1}\langle \hat S_0^{(1)}\rangle
+ \frac{1-\eta^2}{2}\Sigma_{r_1}
+ \eta(1-\eta)\Sigma_{r_2}\langle \hat S_0^{(2)}\rangle
+ \frac{1-\eta^2}{2}\Sigma_{r_2}.
\end{equation}
For simplicity, we consider a state for which \( \frac12\langle \hat q_k\hat p_k + \hat p_k\hat q_k\rangle=\langle \hat q_1\otimes\hat p_2\rangle=\langle \hat p_1\otimes\hat q_2\rangle=0.\) Setting these terms to zero aligns the state with its standard form where conjugate quadratures are decoupled, thereby simplifying the algebra while retaining essential physical features such as entanglement. However, for non-zero values of these entries, our method remains fully applicable.

Finally, the full primed covariance matrix is
\[
\Gamma'(\eta)=
\begin{pmatrix}
\langle \hat q_1'^2\rangle & 0 & \langle \hat q_1'\otimes\hat q_2'\rangle & \langle \hat q_1'\otimes\hat p_2'\rangle \\
0 & \langle \hat p_1'^2\rangle & \langle \hat p_1'\otimes\hat q_2'\rangle & \langle \hat p_1'\otimes\hat p_2'\rangle \\
\langle \hat q_1'\otimes\hat q_2'\rangle & \langle \hat p_1'\otimes\hat q_2'\rangle &
\langle \hat q_2'^2\rangle & 0 \\
\langle \hat q_1'\otimes\hat p_2'\rangle & \langle \hat p_1'\otimes\hat p_2'\rangle & 
0 & \langle \hat p_2'^2\rangle
\end{pmatrix}.
\]

In our analysis, it is important to emphasize that the primed moments obtained after introducing detector loss through lossy beam splitters are not required to coincide with the exact action of the physical loss channel on the covariance matrix. Instead, our objective is different: we wish to evaluate how the full reconstruction protocol behaves when the experimental setup is modified by inserting lossy beam splitters prior to detection. Thus, if we consider only the effect of detection inefficiency here through the corresponding channel action, the aforesaid results will not be reproduced. One should also take into account properly the measurement of the Stokes-like observables as well as the said inversion formulae in order to generate the appropriate channel action. The resulting covariance matrix \(\Gamma'(\eta)\)~therefore represents the output of the reconstruction procedure under realistic detection inefficiencies, rather than the true covariance matrix of the lossy quantum state. This approach allows us to quantify the robustness of the Stokes-based reconstruction method by directly comparing \(\Gamma'(\eta)\) with the ideal covariance matrix \(\Gamma\), thereby identifying regimes of detector efficiency for which entanglement detection remains reliable. This includes evaluating whether the separability criterion given in $\eqref{phys31aa}$  can still detect the entanglement despite the error terms introduced into the elements of the covariance matrix.

While our method is general and applicable to any two- mode Gaussian state, as an example we consider a symmetric two-mode squeezed vacuum (TMSV) input with squeezing parameter $r$,  
the ideal covariance matrix is
\[
\Gamma_{\text{TMSV}}=
\begin{pmatrix}
v\openone{}_2 & \mathrm{diag}(c,-c)\\
\mathrm{diag}(c,-c) & v\openone{}_2
\end{pmatrix},
\qquad
v=\tfrac12 \cosh 2r,\quad c=\tfrac12\sinh 2r.
\]
\(\Gamma'(\eta)\) represents the reconstructed covariance
matrix generated by the
Stokes-like reconstruction method with  displaced-squeezed-thermal (DST) reference states, while explicitly accounting for detection losses.

To avoid divergence in \(1/D_{cd}\), it is necessary to introduce asymmetry between the reference states labeled ``c" and ``d". For the choice of parameters~(See Appendix~\ref{reference})~$\beta_{c} = \pi/4$, $\gamma_{c} = \pi/4$, squeezing parameter $\theta_{c} = 1$, and the average photon number $\overline{n}_{c} = 10^3$, the allowed displacement is $d_{c} = 50.65$. In contrast,  for the reference state ``d", an asymmetry is introduced by setting $\theta_{d} = 0.5$, while keeping all other parameters the same as those of the reference state ``c'', resulting in a displacement $d_{d} = 28.83$. Additionally, total mean number of photons measured at the outputs of the two beam splitters $\text{B.S}_2$ and $\text{B.S}_3$ (see Fig.\ref{figg1}b) are  \(\hat{S}_{0}^{(34)}=n_c^{(t)}+\sinh^2r\), and \(\hat{S}_{0}^{(56)}=n_d^{(t)}+\sinh^2r\), respectively, where \(n_{c(d)}^{(t)}\) denotes total mean number of photons in the reference beam labeled by \(c(d)\), and ``\(\sinh^2{r}\)" is the mean photon number per mode of the arbitrary two-mode Gaussian state (equal in both modes for a two-mode squeezed vacuum state). 
Below, we present the details in a tabular in Table~\ref{table:reference_states_reordered}:


\begin{table}[h!]
\centering
\renewcommand{\arraystretch}{1.3}
\begin{tabular}{c|c|c|c|c}
\hline
\textbf{Parameter} 
& \textbf{Single Mode $r_1$} 
& \textbf{Single Mode $r_2$} 
& \textbf{Ref.\ State ``c''} 
& \textbf{Ref.\ State ``d''} \\ \hline

$\beta$ 
& $\pi/4$ 
& $\pi/4$ 
& $\pi/4$ 
& $\pi/4$ \\

$\gamma$ 
& $\pi/4$ 
& $\pi/4$ 
& $\pi/4$ 
& $\pi/4$ \\

$\theta$ 
& $1$ 
& $1$ 
& $1$ 
& $0.5$ \\

$\overline{n}$ 
& $10^{3}$ 
& $10^{3}$ 
& $10^{3}$ 
& $10^{3}$ \\

$d$ 
& $50.65$ 
& $50.65$ 
& $50.65$ 
& $28.83$ \\ \hline
\end{tabular}
\caption{Parameter choices for the single-mode reference states $r_1$, $r_2$ and the two-mode asymmetric reference states ``c'' and ``d''.}
\label{table:reference_states_reordered}
\end{table}

\end{document}